\documentclass[aps, prd, letterpaper, showkeys, showpacs, twocolumn, nofootinbib, preprintnumbers, longbibliography, unsortedaddress, 10pt]{revtex4-1}

\usepackage[top=2.8cm, bottom=2.8cm, left=2cm, right=2cm]{geometry}
\usepackage[utf8]{inputenc}  
\usepackage{amsmath,amssymb,lmodern,amsfonts} 
\usepackage{graphicx}
\usepackage{natbib}
\usepackage[caption=false]{subfig}
\usepackage{hyperref} 
\usepackage{color}
\usepackage[table]{xcolor}
\usepackage{booktabs}
\usepackage{tabulary}
\usepackage{tabularray}
\usepackage{natbib}
\usepackage{comment}

\definecolor{darkgreen}{rgb}{0.2, 0.5, 0.3}

\begin{document}

\preprint{}

\title{Late-Time Anisotropy Sourced by a 2-Form Field Non-Minimally Coupled to Cold Dark Matter}

\author{J. Bayron Orjuela-Quintana}
\email{john.orjuela@correounivalle.edu.co}
\author{Jose L. Palacios-C\'ordoba}
\email{palacios.jose@correounivalle.edu.co}
\author{C\'esar A. Valenzuela-Toledo}
\email{cesar.valenzuela@correounivalle.edu.co}
\affiliation{Departamento  de  F\'isica,  Universidad  del Valle, \\ Ciudad  Universitaria Mel\'endez,  Santiago de Cali  760032,  Colombia}

\begin{abstract}
This paper investigates the cosmological dynamics arising from the interaction between a 2-form field and cold dark matter within a Bianchi I background. Employing a dynamical system analysis, we identify two attractors yielding to exponential expansion of the Universe, i.e., de-Sitter solutions. Notably, these solutions exhibit a pivotal distinction: one is indistinguishable from the cosmological constant scenario, while the other corresponds to an \emph{anisotropic de-Sitter} expansion sourced by the 2-form field. To validate the asymptotic behavior of our model, we conduct a numerical exploration of its expansion history. Our analysis reveals that the coupling between the dark sectors amplifies the shear during the matter-dominated epoch, offering a potential avenue to address certain observational discrepancies related to the structure formation process. Then, we constrain the parameter space of the model using recent observational datasets. Remarkably, we find that the current shear is precisely constrained to be approximately $\Sigma_0 \approx - 10^{-4}$. We also discuss some key differences in the expansionary dynamics sourced by the 2-form field compared to a 1-form field, i.e., a vector field, offering insights into their respective impacts on the support they provide to late-time anisotropy.
\end{abstract}

\pacs{98.80.Cq, 95.36.+x}

\keywords{late-time anisotropy, interacting dark energy-dark matter, $p$-forms}

\maketitle

\section{Introduction} 
\label{Sec: Intro}

The standard cosmological model is currently facing significant challenges~\cite{Perivolaropoulos:2021jda}. The classical issues concerning the nature of the cosmological constant $\Lambda$~\cite{Weinberg:1988cp} and cold dark matter (CDM) are compounded by observational discrepancies, including the cosmic microwave background (CMB) anomalies~\cite{Bennett:2010jb, Schwarz:2015cma, Gruppuso:2017nap, Muir:2018hjv, Cayuso:2019hen, Shaikh:2019dvb, Billi:2019vvg, Chiocchetta:2020ylv}, the $H_0$~\cite{Planck:2018vyg, Riess:2021jrx, Freedman:2021ahq, Capozziello:2024stm} and $S_8$ tensions~\cite{Heymans:2013fya, KiDS:2020suj, DES:2021bvc, DES:2021vln, Preston:2023uup, Dalal:2023olq, Li:2023tui}, and the cosmic dipole anomaly~\cite{Secrest:2020has, Dalang:2021ruy, Luongo:2021nqh, Dam:2022wwh, Mastrogiovanni:2022nya, Cheng:2023eba}, among others~\cite{Fields:2011zzb, Cyburt:2015mya, Bowman:2018yin, Pitrou:2020etk, Singh:2021mxo, Abdalla:2022yfr}. These inconsistencies necessitate a thorough reassessment of the underlying assumptions of the $\Lambda$CDM paradigm, particularly the cosmological principle which posits that the Universe's expansion is statistically homogeneous and isotropic at cosmological scales. The validity of this principle is increasingly questioned~\cite{Colin:2019opb, Aluri:2022hzs, Hu:2023eyf, Jones:2023ncn}, stimulating investigations into viable alternatives to $\Lambda$CDM.

In this context, we explore three extensions to the standard theoretical paradigm. Firstly, we consider replacing $\Lambda$ with a dynamical dark energy (DE) component, which could potentially resolve the naturalness issues associated with interpreting $\Lambda$ as vacuum energy~\cite{Martin:2012bt}. Secondly, we investigate scenarios featuring anisotropic late-time accelerated expansion, which could provide insights into observed anomalies that suggest violations of rotational invariance~\cite{Battye:2009ze, Akarsu:2009bhh, Campanelli:2011uc}. Thirdly, we explore the possibility of interactions between DE and CDM, which are not precluded by any known symmetry and may significantly influence cosmological dynamics~\cite{Bolotin:2013jpa, Wang:2016lxa}. Then, our study focuses on an interacting CDM-DE model within an anisotropic expansion framework.

Concerning the relaxation of the isotropy assumption, an anisotropic DE component could leave observable imprints on various cosmological phenomena~\cite{Fleury:2014rea}. For instance, they affect the lowest multipoles of the CMB angular power spectrum~\cite{Campanelli:2006vb, BeltranJimenez:2007rsj, Marcori:2016oyn}, the structure formation process~\cite{Koivisto:2005mm, Dimastrogiovanni:2008ua, BeltranAlmeida:2021ywl}, and could be relevant in weak lensing data analyses~\cite{Pitrou:2015iya}. Such effects can be instrumental in constraining the permissible levels of anisotropy at present~\cite{Schucker:2014wca, Hossienkhani:2019kew, Akarsu:2021max}. Typical tests of late-time isotropy, primarily based on supernovae measurements, provide local bounds on the current shear magnitude $|\Sigma_0| < \mathcal{O}(0.001)$ at a 68\% confidence level~\cite{Campanelli:2011uc, Amirhashchi:2018nxl}. In contrast, CMB analyses impose far stricter constraints, approximately $10^{-10}$~\cite{Saadeh:2016sak}.  However, these bounds are subject to reliability issues when considering anisotropy sourced by a DE component. Indeed, the large-scale isotropy assumption is an open issue yet~\cite{Kolatt:2000yg, Appleby:2012as, BeltranJimenez:2014otq, Bengaly:2015dza, Bengaly:2015xkw, Andrade:2019kvl, Rahman:2021mti}.

From a theoretical point of view, the cosmic no-hair conjecture postulates that any expanding Universe will asymptotically evolve towards an isotropic de Sitter space~\cite{Hawking:1981fz, Barrow:1987ia}. Long ago, Wald proved that this conjecture holds in all Bianchi geometries considering a positive cosmological constant and that all matter fluids comply with the dominant and strong energy conditions~\cite{Wald:1983ky}. However, violations of these conditions could potentially circumvent this conjecture. Although scalar fields are commonly employed in DE scenarios, they do not naturally sustain prolonged anisotropic expansion. This limitation has been addressed by introducing couplings to fields that inherently break isotropy, such as vector fields~\cite{Thorsrud:2012mu, Orjuela-Quintana:2021zoe, Garcia-Serna:2023xfw, Gallego:2024gay} and 2-form fields~\cite{BeltranAlmeida:2019fou}. Other options involve modified gravity theories~\cite{Akarsu:2013dva, Akarsu:2019pvi}, non-abelian gauge fields~\cite{Orjuela-Quintana:2020klr, Guarnizo:2020pkj}, and inhomogeneous scalar fields~\cite{Motoa-Manzano:2020mwe, BeltranAlmeida:2024uvk}.

In terms of the economy principle, it would be interesting to identify a model where a single field drives an \emph{anisotropic accelerated expansion} phase. Initially, vector fields present a promising avenue. However, investigations within the realm of beyond-generalized Proca theories reveal significant limitations. As demonstrated in Ref.~\cite{Heisenberg:2016wtr}, vector fields drive the accelerated expansion towards an isotropic de Sitter configuration, aligning with the predictions of the cosmic no-hair conjecture.  This convergence towards isotropy necessitates the exploration of alternative candidates. Consequently, 2-form fields emerge as the next logical choice for potentially realizing anisotropic accelerated expansion. 

In Ref.~\cite{Kaloper:1991rw}, it has been  shown that a 2-form field, in conjunction with a positive cosmological constant, can deviate from the cosmic no-hair conjecture in certain Bianchi geometries. Previous works have also explored the role of the 2-form field as a principal contributor to anisotropy when coupled with other fields that drive accelerated expansion (see Refs.~\cite{BeltranAlmeida:2018nin, Almeida:2019xzt, Normann:2019rgo, Normann:2019ceb, Almeida:2020lsn}). Yet, the potential for a 2-form field to solely induce an anisotropic accelerated expansion remains unexplored. $p$-forms, which are prevalent in higher-dimensional theories such as string theory, manifest in the low-energy effective action of gravity. In this setup, the kinetic term associated with the Kalb-Ramond 2-form field $B_{\mu\nu}$ is incorporated into the gravitational action. In its vacuum state, this action includes a cosmological constant, reflecting the approach used in Kaloper's work~\cite{Kaloper:1991rw}. However, after the quantum gravity phase, the Kalb-Ramond field may acquire an effective potential, thereby influencing subsequent cosmological epochs.

On the other hand, although an anisotropic accelerated expansion could address several cosmological challenges, including the noted $H_0$ tension~\cite{Amirhashchi:2020qep, Yadav:2023yyb}, some discrepancies during the structure formation process still lack a solution. It has been shown that a momentum transfer between dark sectors could ameliorate the $S_8$ tension~\cite{Cardona:2022mdq}. Such interactions are not only plausible but might be necessary for a consistent theory, barring any unknown symmetries that prevent energy and momentum exchanges between these components.

Summarizing, in this paper, we investigate the cosmological dynamics driven by the Kalb-Ramond 2-form field, which we assume is not in its vacuum state; thus, a potential term is included in its formulation. This investigation is set on top of a Bianchi background, with the 2-form field allowed to interact with CDM, modeled as a pressure-less fluid. This paper is organized in the following way. The general framework and the corresponding equations of motion for the fields in the anisotropic background are presented in Sec.~\ref{Sec: 2-Form Field Interacting with CDM}. In Sec.~\ref{Sec: Dynamical System}, we use the dynamical system technique to analyze the asymptotic behavior of the model. The cosmological evolution of the fields, for a particular set of parameters and initial conditions, and some of the cosmological implications driven by the interplay between the anisotropy and the fields are presented in Sec.~\ref{Sec: Cosmo Evo}. In Sec.~\ref{Sec: Difference wrt a Vector Field} we present some arguments to highlight some differences in the cosmological evolution of a 1-form and a 2-form field. Then, observational constraints on the parameters of our model are presented in Sec.~\ref{Sec: Cosmological Constraints}. Finally, our conclusions are presented in Sec.~\ref{Sec: Conclusions}.

\section{Dynamics of a 2-Form Field Interacting with CDM}
\label{Sec: 2-Form Field Interacting with CDM}

\subsection{General Framework}

Let us consider a scenario where a 2-form field interacts minimally with gravity but has a non-minimal coupling with CDM. The dynamics of this system are described by the following action:\footnote{Greek indices run from 0 to 3 and denote space-time components, and Latin indices run from 1 to 3 and denote spatial components.}
\begin{align} 
\label{Eq: Action}
S \equiv \int \text{d}^4 x \sqrt{-g} \bigg[ &\frac{m_\text{P}^2}{2} R - \frac{1}{12} H_{\mu\nu\rho} H^{\mu\nu\rho} - V(X) \nonumber \\
 &+ f(X) \mathcal{L}_c + \mathcal{L}_b + \mathcal{L}_r \bigg].
\end{align}
Here, $g$ denotes the determinant of the metric, $m_\text{P}$ represents the reduced Planck mass, $R$ is the Ricci scalar, and $H_{\mu\nu\rho}$ is the field strength tensor associated with the 2-form field $B_{\mu\nu}$, which is formally defined as
\begin{equation}
\label{Eq: Strength Tensor}
    H_{\mu\nu\rho} \equiv 3 \nabla_{[\mu} B_{\nu\rho]} = \nabla_\mu B_{\nu\rho} + \nabla_\nu B_{\rho\mu} + \nabla_\rho B_{\mu\nu},
\end{equation}
where the square brackets defined standard antisymmetrization. The potential of the 2-form field is denoted as $V(X)$, where $X \equiv B_{\mu\nu} B^{\mu\nu}$. Additionally, $f(X)$ represents the non-minimal coupling function between the dark sectors, and $\mathcal{L}_c$, $\mathcal{L}_b$, and $\mathcal{L}_r$ are the Lagrangian densities for CDM, baryonic matter, and radiation, respectively. 

Varying the action in Eq.~\eqref{Eq: Action} with respect to the metric yields the gravitational field equations:
\begin{equation} 
\label{Eq: Einstein Eqs}
    m_\text{P}^2 G_{\mu\nu} = T_{\mu\nu},
\end{equation}
where $G_{\mu\nu}$ is the Einstein tensor, and $T_{\mu\nu}$ is the total energy tensor given by:
\begin{equation}
    T_{\mu\nu} \equiv T^{(B)}_{\mu\nu} + f(X) T^\text{(c)}_{\mu\nu} + T^{(b)}_{\mu\nu} + T^{(r)}_{\mu\nu},
\end{equation}
where $T^\text{(c)}_{\mu\nu}$, $T^{(b)}_{\mu\nu}$, and $T^{(r)}_{\mu\nu}$ represent the energy tensors of the CDM, baryonic, and radiation fluids, respectively. The contribution of the 2-form field corresponds to the following energy tensor:
\begin{align} 
\label{Eq: 2-form Energy Tensor}
    T^{(B)}_{\mu\nu}  &\equiv \frac{1}{2} H_{\mu\rho\sigma} H_{\nu}^{\ \rho\sigma} - g_{\mu\nu} \left[ \frac{1}{12} H_{\lambda\rho\sigma} H^{\lambda\rho\sigma} + V(X) \right] \nonumber \\
    &+  4 \left( V_X - \frac{f_X}{f} \tilde{\mathcal{L}}_\text{c} \right) B_{\mu\rho} B_\nu^{\ \rho}.
\end{align}
Here, we have defined $\tilde{\mathcal{L}}_c \equiv f(X) \mathcal{L}_c$, and utilized the shorthand notation $V_X \equiv \tfrac{\text{d} V}{\text{d} X}$ and $f_X \equiv \tfrac{\text{d} f}{\text{d} X}$. Although we have separated the energy tensor contributions from each fluid, it is important to realize that this division is somewhat arbitrary. Note, for instance, that the Lagrangian of CDM contributes to $T^{(B)}_{\mu\nu}$ in the latter equation. We will explore this aspect further in the subsequent sections.

Finally, upon variation of the action in Eq.~\eqref{Eq: Action} with respect to $B_{\mu\nu}$, we derive the equation of motion that governs the dynamics of the coupled 2-form sector:
\begin{equation}
\label{Eq: 2-form EoM}
    \nabla_\rho H^{\mu\nu\rho} = 4 B^{\mu\nu} \left( V_X -  \frac{f_X}{f} \tilde{\mathcal{L}}_c \right).
\end{equation}

In the upcoming section, we will delve into the dynamics of this model within a cosmological framework.

\subsection{Dynamics in a Bianchi I Cosmology}

As highlighted in the Introduction section, certain observational clues suggest a departure from the cosmological principle as an accurate description of our Universe on cosmological scales~\cite{Aluri:2022hzs}. Consequently, it becomes imperative to move beyond the Friedman-Lema\^itre-Robertson-Walker (FLRW) metric as the fundamental object describing the geometry of our Universe. Instead, we need to entertain more general geometries that encompass both inhomogeneities and anisotropy. Since the most statistically significant observational anomalies point to a violation in the isotropy on large scale~\cite{Schwarz:2015cma, Jones:2023ncn}, we concentrate on homogeneous but anisotropic geometries, which are naturally accommodated into the realm of Bianchi cosmologies. Among these, the simplest realization is the Bianchi I metric, expressed in Cartesian coordinates as:
\begin{equation}
\label{Eq: Bianchi I Metric}
    \text{d} s^2 = - \text{d} t^2 + a(t)^2 e^{2\beta_i (t)} \delta_{ij} \text{d} x^i \text{d} x^j,
\end{equation}
where the expansion of the Universe is described by an average scale factor $a(t)$, as a function of the cosmic time $t$, while the anisotropy is encoded within the functions $\beta_i$. These functions are subject to the volume constraint:
\begin{equation}
    \sum_{i = 1}^3 \beta_i(t) = 0.
\end{equation}
It is crucial to emphasize that $\beta_i$ do not correspond to the components of a vector field. Therefore, the Einstein summation rule does not apply in this context~\cite{Pereira:2007yy}. 

Our main objective entails elucidating anisotropic accelerated solutions driven by a self-interacting 2-form field, along with potential alterations stemming from non-minimal coupling with CDM. To achieve this, we opt for a further simplification of the Bianchi-I metric. As a preliminary exploration, we adopt the concept of \emph{local rotational symmetry} (LRS) within the geometry. Under LRS, the anisotropy manifests through a single function, termed the geometric shear $\sigma(t)$. Then, the anisotropy functions $\beta_i$ can be expressed as follows:
\begin{equation}
\label{Eq: LRS Condition}
    \beta_1 (t) \equiv -2\sigma(t), \quad \beta_2 (t) = \beta_3 (t) \equiv \sigma (t).
\end{equation}
The profile of the 2-form has to be chosen accordingly to preserve the symmetries of the background geometry. In this instance, the homogeneous 2-form field adheres to rotational symmetry within the $(y, z)$ plane. This configuration can be expressed as:
\begin{equation} 
\label{Eq: 2-form Ansatz}
    B_{\mu\nu} \, \text{d} x^\mu \wedge \text{d} x^\nu \equiv 2 \phi (t) \, \text{d} y \wedge \text{d} z,
\end{equation}
being $B_{23} = - B_{32} = \phi(t)$ the only non-vanishing component, and $\phi(t)$ a scalar field. This particular choice for the field profile is not arbitrary; as noted in Ref.~\cite{Normann:2017aav}, within a LRS cosmology, the isotropy-violating field must be aligned with the Killing vector field that represents invariance under spatial rotation. Therefore, in this scenario, the 2-form field must be ``aligned'' with the $x$-axis of the Bianchi metric.

By applying the LRS Bianchi-I metric from Eqs.~\eqref{Eq: Bianchi I Metric} and \eqref{Eq: LRS Condition}, along with the assumption outlined for the 2-form field in Eq.~\eqref{Eq: 2-form Ansatz}, we determine that only certain components of the strength tensor $H_{\mu\nu\rho}$, defined in Eq.~\eqref{Eq: Strength Tensor}, are non-zero. Specifically, we find:
\begin{equation}
    H_{023} = \dot{\phi},
\end{equation}
and the corresponding components obtained by rearranging the indices. Accordingly, the terms representing the field's kinetic energy, denoted as  $Y \equiv H_{\mu\nu\rho} H^{\mu\nu\rho}$, and the quadratic term, $X \equiv B_{\mu\nu} B^{\mu\nu}$, can be expressed as:
\begin{equation}
\label{Eq: Quadratic Terms}
    X = 2 \frac{\phi^2 e^{-4 \sigma}}{a^4}, \quad Y = - 6 \frac{\dot{\phi}^2 e^{-4 \sigma}}{a^4},
\end{equation}

The energy tensor associated to the 2-form field, described in Eq.~\eqref{Eq: 2-form Energy Tensor}, yields the following non-zero components:
\begin{align}
    T^{0 (B)}_{\ 0}&= - \frac{1}{2} \frac{\dot{\phi}^2 e^{-4 \sigma}}{a^4} - V, \label{Eq: T00} \\
    T^{1 (B)}_{\ 1} &= \frac{1}{2} \frac{\dot{\phi}^2 e^{-4 \sigma}}{a^4} - V, \label{Eq: T11} \\
    T^{2 (B)}_{\ 2} &= - \frac{1}{2} \frac{\dot{\phi}^2 e^{-4 \sigma}}{a^4} - V, \label{Eq: T22} \\
    & \quad \, + 4 \left( V_X - \frac{f_X}{f} \tilde{\mathcal{L}}_c\right)\frac{\phi^2 e^{-4\sigma}}{a^4}, \nonumber \\
    T^{3 (B)}_{\ 3} &= T^{2 (B)}_{\ 2}.
\end{align}
The expansion history of the Universe is dictated by the evolution of the average scale factor coupled with the cosmic fluids. This evolution is governed by the Friedman equations and the equation of motion of the 2-form field. Taking the ``00'' component of the field equations [Eqs.~\eqref{Eq: Einstein Eqs} and using Eq.~\eqref{Eq: T00}], we derive the first Friedman equation:
\begin{align}
\label{Eq: Friedman 1 Eq}
    3 m_\text{P}^2 H^2 &= \frac{1}{2} \frac{\dot{\phi}^2 e^{-4 \sigma}}{a^4} + V + \rho_r + \rho_b + f(X)\rho_c \\
    &+ 3 m_\text{P}^2 \dot{\sigma}^2, \nonumber
\end{align}
where $\rho_c$, $\rho_b$, $\rho_r$ are the densities of CDM, baryons, and radiation, respectively. Assuming that baryonic and CDM are pressure-less fluids, the trace of the Einstein equations ($m_\text{P}^2 G^{i}_{\ i} = T^i_{\ i}$) provides the second Friedman equation:
\begin{align}
    - 2 m_\text{P}^2 \dot{H} &= \frac{1}{3} \frac{\dot{\phi}^2 e^{-4\sigma}}{a^4} + \frac{8}{3} \left( V_X - \frac{f_X}{f} \tilde{\mathcal{L}}_c \right) \frac{\phi^2 e^{-4 \sigma}}{a^4} \nonumber \\
    &+ f(X) \rho_c + \rho_b + \frac{4}{3} \rho_r + 6 m_\text{P}^2 \dot{\sigma}^2. \label{Eq: Friedman 2 Eq}
\end{align}
The equation governing the evolution of the geometrical shear is obtained from the linear combination \mbox{$m_\text{P}^2 \left( G^2_{\ 2} - G^1_{\ 1} \right) = (T^2_{\ 2} - T^1_{\ 1})$} as:
\begin{align} 
    3 m_\text{P}^2 \ddot{\sigma} + 9 m_\text{P}^2 H \dot{\sigma} = &- \frac{\dot{\phi}^2 e^{-4\sigma}}{a^4} \nonumber \\
    &+ 4 \left( V_X - \frac{f_X}{f} \tilde{\mathcal{L}}_c \right) \frac{\phi^2 e^{-4 \sigma}}{a^4}. \label{Eq: Geometric Shear}
\end{align}
Finally, from Eq.\eqref{Eq: 2-form EoM}, with the ansatz in Eq.~\eqref{Eq: 2-form Ansatz}, we derive the equation of motion for the only dynamical degree of freedom of the 2-form field:
\begin{equation} 
\label{Eq: phi EoM}
    \ddot{\phi} - (H + 4 \dot{\sigma}) \dot{\phi} + 4 \left( V_X - \frac{f_X}{f} \tilde{\mathcal{L}}_c \right) \phi = 0.
\end{equation}

Given our focus on describing the late-time accelerated expansion of the Universe, it is crucial to distinguish all the terms contributing to the dark energy fluid. This can be done by looking at all the terms in the first Friedman equation, which in general can be written as
\begin{equation}
    3 m_\text{P}^2 H^2 = \sum_i \rho_i,
\end{equation}
where $i = c, b, r, \text{DE}$ stands for all the matter fluids in the cosmic inventory. We identify the density of the interacting CDM as:
\begin{equation}
\label{Eq: Density Interacting CDM}
    \tilde{\rho}_c \equiv f(X) \rho_c,
\end{equation}
whereas the density of dark energy will take the form
\begin{equation} 
\label{Eq: DE Density}
    \rho_\text{DE} \equiv \frac{1}{2} \frac{\dot{\phi}^2 e^{-4 \sigma}}{a^4} + V + 3 m_\text{P}^2 \dot{\sigma}^2.
\end{equation}

While the interaction between the dark components is encapsulated in the Lagrangian density $\tilde{\mathcal{L}}_c$, the true interaction kernel becomes explicit in the continuity equations for the interacting fluids. In a general context, we expect
\begin{align} 
    \dot{\tilde{\rho}}_c + 3 H \tilde{\rho}_c &= \mathcal{Q}, \label{Eq: Continuity CDM} \\
    \dot{\rho}_\text{DE} + 3 H \left( \rho_\text{DE} + p_\text{DE} \right)  &= - \mathcal{Q} \label{Eq: Continuity DE},
\end{align}
where $p_\text{DE}$ denotes the pressure of dark energy, and $\mathcal{Q}$ represents the interaction kernel. Thus, when $\mathcal{Q} > 0$, DE transitions into CDM, while $\mathcal{Q} < 0$ indicates CDM transitioning into DE~\cite{Wang:2016lxa}. In our scenario, we refrain from imposing any condition on the sign of $\mathcal{Q}$.\footnote{However, thermodynamic considerations suggest that dark energy must decay into dark matter~\cite{Pavon:2007gt}.}

By computing the time derivative of the density in Eq.~\eqref{Eq: Density Interacting CDM}, the continuity equation for the interacting CDM fluid reads
\begin{equation} 
\label{Eq: Continuity Interacting CDM}
    \dot{\tilde{\rho}}_c + 3 H \tilde{\rho}_c = 2 \frac{\text{d} \ln f}{\text{d} \ln X} \left[ \frac{\dot{\phi}}{\phi} - 2 \left( H + \dot{\sigma} \right) \right] \tilde{\rho}_c,
\end{equation}
where we account for $\dot{f} = f_X \dot{X}$, with $X$ specified in Eq.~\eqref{Eq: Quadratic Terms}, and the uncoupled CDM follows the standard continuity equation $\dot{\rho}_c = - 3 H \rho_c$. Then, computing the time derivative of the DE density defined in Eq.~\eqref{Eq: DE Density}, we obtain:
\begin{equation} \label{continuity DE}
    \dot{\rho}_\text{DE} + 3 H \left( \rho_\text{DE} + p_\text{DE} \right) = 2 \frac{\text{d} \ln f}{\text{d} \ln X} \left[ \frac{\dot{\phi}}{\phi} - 2\left( H + \sigma \right) \right] \tilde{\mathcal{L}}_c,
\end{equation}
where we have used Eq.~\eqref{Eq: Geometric Shear} to eliminate $\ddot{\sigma}$, and Eq.~\eqref{Eq: phi EoM} to eliminate $\ddot{\phi}$. To derive the latter equation, we identify the DE pressure with:
\begin{align} 
\label{Eq: DE Pressure}
    p_\text{DE} &\equiv - \frac{1}{6} \frac{\dot{\phi}^2 e^{-4 \sigma}}{a^4} - V \nonumber \\
    &+ \frac{8}{3} \left( V_X - \frac{f_X}{f} \tilde{\mathcal{L}}_c \right) \frac{\phi^2 e^{-4\sigma}}{a^4} + 3 m_\text{P}^2 \dot{\sigma}^2.
\end{align}
This equation reveals that the interaction term introduces an effective pressure for the 2-form field, resembling the scenario of a vector field coupled to CDM~\cite{Gomez:2020sfz}.

From Eqs.~\eqref{Eq: Continuity CDM} and \eqref{Eq: Continuity DE}, it is clear that the interaction kernel $\mathcal{Q}$ has the form:
\begin{equation} 
\label{Eq: Interaction Kernel}
    \mathcal{Q} \equiv  2 \frac{\text{d} \ln f}{\text{d} \ln X} \left[ \frac{\dot{\phi}}{\phi} - 2\left( H + \sigma \right) \right] \tilde{\rho}_c,
\end{equation}
which requires that:
\begin{equation}
    \tilde{\mathcal{L}}_c = - \tilde{\rho}_c.
\end{equation}
Here, we wish to highlight two key points. First, the identification of $\tilde{\mathcal{L}}_c = - \tilde{\rho}_c$ aligns with expectations from the behavior of an ideal fluid~\cite{Schutz:1970my, Brown:1992kc, Bertolami:2008ab, Mendoza:2020bzc}, serving as a foundational requirement of the theory. This is also a consequence of the definition of the density and pressure of dark energy incorporating contributions from the geometrical shear [see Eqs.~\eqref{Eq: DE Density} and \eqref{Eq: DE Pressure}]. Although these definitions might appear unconventional, they are necessitated by the theory to yield the standard continuity equations.\footnote{When the anisotropy is not taken into account in the definition of the DE density, its continuity equation is given by \mbox{$\dot{\rho}_\text{DE} + 3H(\rho_\text{DE} + p_\text{DE}) \propto \dot{g}_{ij}\Pi^{i j}$}, where $g_{ij}$ is the spatial part of the metric, and $\Pi^{i j}$ is the trace-free part of the energy-momentum tensor. In that case, the identification of $\mathcal{Q}$ would not be as easier as done here.} Second, concerning the interaction kernel, particle physics reasoning suggests that $\mathcal{Q}$ should be expressed in terms of the density of one or both interacting fluids and a function of time. In the absence of a fundamental theory providing explicit formulations for $\mathcal{Q}$, this expectation has inspired the investigation of phenomenological kernels such as $\mathcal{Q} \propto H \rho_c$~\cite{Valiviita:2008iv, Boehmer:2008av}, or $\mathcal{Q} \propto H (\rho_c + \rho_\text{DE})$~\cite{Zimdahl:2001ar}, among others~\cite{Boehmer:2009tk}. In this context, our theory offers a more grounded proposal for $\mathcal{Q}$ from a field-theoretical perspective, even though the coupling function $f(X)$ remains to be specified.

To unveil the cosmological background dynamics of our model, the set of Eqs.~\eqref{Eq: Friedman 1 Eq}-\eqref{Eq: phi EoM} are supplemented with the continuity equations for interacting fluids, Eqs.~\eqref{Eq: Continuity CDM} and \eqref{Eq: Continuity DE}, and the standard continuity equations for baryons and radiation. However, significant insights can be gleaned by analyzing the asymptotic evolution of the model. In the next section, we will study the asymptotic behavior of this set of equations through a dynamical system analysis.

\section{Dynamical System}
\label{Sec: Dynamical System}

\subsection{Autonomous Set}

In this section, we employ the well-established technique of dynamical systems to elucidate the asymptotic cosmological behavior inherent in our model~\cite{Coley:2003mj, Wainwright2009, Bahamonde:2017ize}. To achieve this, we transition from the set of differential equations delineated in Eqs.~\eqref{Eq: Friedman 1 Eq}-\eqref{Eq: phi EoM} to an autonomous set of first-order differential equations.

Beginning with Eq.~\eqref{Eq: Friedman 1 Eq}, we introduce the following dimensionless variables:
\begin{equation*}
    z \equiv \frac{1}{\sqrt{6} m_\text{P}} \frac{\dot{\phi} e^{-2 \sigma}}{a^2 H}, \quad v \equiv \frac{1}{m_\text{P} H} \sqrt{\frac{V}{3}}, \quad \Sigma \equiv \frac{\dot{\sigma}}{H},
\end{equation*}
\begin{equation}
\label{Eq: Variables}
    \tilde{\Omega}_c \equiv \frac{\tilde{\rho}_c}{3 m_\text{P}^2 H^2}, \quad \Omega_b \equiv \frac{\rho_b}{3 m_\text{P}^2 H^2}, \quad \Omega_r \equiv \frac{\rho_r}{3 m_\text{P}^2 H^2}.
\end{equation}
This transforms the first Friedman equation into the constraint:
\begin{equation} \label{Friedman constraint}
    1 = z^2 + v^2 + \Sigma^2 + \tilde{\Omega}_c + \Omega_b + \Omega_r.
\end{equation}
Utilizing this constraint enables us to eliminate one redundant variable in favor of the others.

Moving to the second Friedman equation in Eq.~\eqref{Eq: Friedman 2 Eq}, we express the deceleration parameter, $q \equiv - 1 - \dot{H}/H^2$, in terms of the dynamical variables, yielding:
\begin{align}
    q &= \frac{1}{2}\left[ 1 - z^2 - 3v^2 + \Omega_r + 3 \Sigma^2 \right] \nonumber \\
      &+ 4\left( v^2 \frac{V_X}{V} + \tilde{\Omega}_c \frac{f_X}{f} \right) \frac{\phi^2 e^{-2\sigma}}{a^4}.
\end{align}
To fully render the system autonomous, additional dimensionless variables are required. Accordingly, we introduce the following variable 
\begin{equation}
\label{Eq: Variable x}
    x \equiv  \frac{\phi e^{-2 \sigma}}{m_\text{P} a^2}.
\end{equation}
In addition, it is necessary to establish the potential and the coupling function. Here, we opt for exponential functions:
\begin{equation}
\label{Eq: Potential and Coupling}
    V(X) \equiv V_0 \, e^{- \lambda X / m_\text{P}^2}, \quad f(X) \equiv f_0 \, e^{ \mu X / m_\text{P}^2},
\end{equation}
where $\lambda > 0$ and $\mu$ are dimensionless constants. Despite the absence of definitive observational constraints on the precise form of these functions, the choice of exponential potentials and couplings is not only motivated for their simplicity but also aligns with conjectures arising from string compactifications, offering a compelling starting point for exploring dark energy mechanisms~\cite{Gallego:2024gay}. With these selections, the deceleration parameter simplifies to:
\begin{equation}
    q = \frac{1}{2} \left[ 1 - z^2 - 3 v^2 - 8 \left(\lambda v^2 - \mu \tilde{\Omega}_c \right) x^2 + \Omega_r + 3 \Sigma^2 \right].
\end{equation}

Using the dimensionless variables defined in Eqs.~\eqref{Eq: Variables} and \eqref{Eq: Variable x}, along with the exponetial potential and coupling function in Eq.~\eqref{Eq: Potential and Coupling}, the set of background equations of motion [Eqs.~\eqref{Eq: Friedman 1 Eq}-\eqref{Eq: phi EoM}] is replaced by the following autonomous set
\begin{align}
    z' &= z \left( q + 2 \Sigma \right) + 2 \sqrt{6} \left( \lambda v^2 - \mu \tilde{\Omega}_c \right) x, \label{Eq: z Eq}\\
    v' &= v(q + 1) - 2 \lambda v x \left[ \sqrt{6} z - 2 x (1 + \Sigma) \right], \\
    x' &= \sqrt{6} z - 2 x (1 + \Sigma), \\
    \Sigma' &= \Sigma (q - 2) - 2 z^2 - 4 \left(\lambda v^2 - \mu \tilde{\Omega}_c \right) x^2, \\
    \Omega_b' &= 2 \Omega_b \left( q - 1/2 \right), \\
    \Omega_r' &= 2 \Omega_r \left( q - 1 \right), \label{Eq: r Eq}
\end{align} 
where a prime represents a derivative with respect to the number of e-folds $N \equiv \ln a$.

To facilitate the identification of late-time accelerated solutions, we characterize the dark energy fluid using its density parameter $\Omega_\text{DE} \equiv \rho_\text{DE} / 3 m_\text{P}^2 H^2$ and its equation of state $w_\text{DE} \equiv p_\text{DE} / \rho_\text{DE}$, expressed in terms of the dimensionless variables as:
\begin{equation}
    \Omega_\text{DE} = z^2 + v^2 + \Sigma^2,
\end{equation}
\begin{equation}
    w_\text{DE} = -1 + \frac{2}{3} \frac{z^2 - 4 \left(\lambda v^2 - \mu \tilde{\Omega}_\text{CDM} \right) x^2 + 3 \Sigma^2}{z^2 + v^2 + \Sigma^2}.
\end{equation}

\subsection{Relevant Fixed Points}

In general, the deceleration parameter delineates the evolution of the average scale factor $a(t)$. Alternatively, we can employ the effective equation of state, $w_\text{eff}$, given by:
\begin{equation} 
\label{Eq: weff}
    w_\text{eff} = - 1 - \frac{2}{3} \frac{\dot{H}}{H^2} \frac{1}{1 - \Sigma^2} - 2 \frac{\Sigma^2}{1 - \Sigma^2},
\end{equation}
which correlates with the deceleration parameter. Since observations imply $\Sigma \ll 1$, then $w_\text{eff} \approx (2q - 1)/3$. This leads to straightforward identifications: $q = 1 \Rightarrow w_\text{eff} \approx 1/3$, $q = 1/2 \Rightarrow w_\text{eff} \approx 0$, and $q < 0 \Rightarrow w_\text{eff} < -1/3$, denoting epochs dominated by radiation, matter, and dark energy, respectively.

In the following, we discuss the fixed points relevant to the radiation era ($\Omega_r \simeq 1, w_\text{eff} \simeq 1/3$), the  matter era ($\Omega_b + \Omega_\text{CDM} \simeq 1, w_\text{eff} \simeq 0$), and the dark energy era ($\Omega_\text{DE} \simeq 1, w_\text{eff} < -1/3$). These fixed points are derived by setting time derivatives of relevant variables to zero, i.e., $z' = 0$, $v' = 0$, $x' = 0$, $\Sigma' = 0$, $\Omega_b' = 0$, and $\Omega_r' = 0$, in the set of equations given in Eqs.~\eqref{Eq: z Eq}-\eqref{Eq: r Eq}, and solving the resulting algebraic equations.  The stability of these fixed points is determined by perturbing the dynamical system around them and analyzing the eigenvalues of the resulting Jacobian matrix. Up to linear order, the perturbations \mbox{$\delta \mathcal{X} = (\delta z, \delta v, \delta x, \delta \Sigma, \delta \Omega_b, \delta \Omega_r)$} satisfy the differential equation,
\begin{equation}
    \delta \mathcal{X} = \mathbb{J} \, \mathcal{X}, 
\end{equation} 
where $\mathbb{J}$ is a $6 \times 6$ Jacobian matrix. The sign of the real part of the eigenvalues $\nu_{1, 2, 3, 4, 5, 6}$ of $\mathbb{J}$ determines the stability of the point. A fixed point is an attractor if all eigenvalues have negative real parts, it is a saddle if at least one eigenvalue is positive while others are negative, and it is a repeller if all eigenvalues are positive.

\subsubsection{\textbf{Radiation Dominance}}

\begin{itemize}
\item[$\bullet$] \textbf{(R1) Isotropic radiation:}
\end{itemize}
This marks the standard isotropic radiation dominated point where
\begin{equation}
    \Omega_r = 1, 
\end{equation}
with all other variables being zero. In this case, the Universe is predominantly filled with ultra-relativistic matter ($w_\text{eff} \sim 1/3$). The eigenvalues of $\mathbb{J}$ at this point are
\begin{equation}
    \nu_{1,\ldots,6} = \{-2, 2, -1, 1, 1, 1\},
\end{equation}
thus rendering it a saddle point with four positive and two negative eigenvalues.

We have discovered another fixed point where $w_\text{eff} = 1/3$. However, the dark energy contribution to the cosmic budget is $\Omega_\text{DE} = 1/2$. We identify this point as non-viable and exclude it from subsequent analysis. Hence, our focus lies on cosmological dynamics originating from the vicinity of point (R1).

\subsubsection{\textbf{Matter Dominance}}

\begin{itemize}
\item[$\bullet$] \textbf{(M1) Isotropic matter:}
\end{itemize}

This represents the conventional isotropic matter point where baryonic matter and CDM dominate the Universe's energy content ($w_\text{eff} \sim 0$):
\begin{equation}
    \Omega_b + \tilde{\Omega}_c = 1, 
\end{equation}
with all other variables being zero. The eigenvalues of $\mathbb{J}$ are
\begin{align}
    \nu_{1, \ldots, 4} &= \left\{ \frac{3}{2}, -\frac{3}{2},  -1, 0 \right\}, \\
    \nu_{5, 6} &= \frac{3}{4} \left[ - 1 \pm \sqrt{25 - 192 \mu (1 - \Omega_b)} \right].
\end{align}
Although non-hyperbolic, i.e., at least one eigenvalue of the Jacobian matrix at the point is zero this point acts as a saddle due to a mix of positive and negative eigenvalues. 

\begin{itemize}
\item[$\bullet$] \textbf{(M2) Scaling CDM-DE:}
\end{itemize}

Here, we identify a point where CDM and DE coexist. The variables are expressed as follows:
\begin{equation*}
    z = \frac{\sqrt{6} x}{3 + 2 x^2}, \quad \Sigma = - \frac{2 x^2}{3 + 2 x^2}, \quad \Omega_\text{DE} = \frac{2 x^2}{3 + 2x^2},
\end{equation*}
\begin{equation}
\label{Eq: M2}
    \Omega_b = \frac{3(12\mu -1) + 6x^2(1 + 4\mu)}{4\mu(3 + 3x^2)^2},
\end{equation}
where
\begin{equation} 
    \tilde{\Omega}_c = 1 - \Omega_b - \Omega_\text{DE} = \frac{1}{4\mu} \frac{3 - 6 x^2}{(3 + 2x^2)^2},
\end{equation}
and $w_\text{DE} = 0$, i.e., dark energy behaves as a pressure-less fluid. This critical manifold signifies a scaling solution between the dark components, alongside the presence of baryons, as described by the simple relation
\begin{equation}
    \frac{\Omega_\text{DE}}{\tilde{\Omega}_c + \Omega_b} = \frac{2}{3}x^2.
\end{equation}
However, this point is deemed non-viable as well. Notably, Eq.~\eqref{Eq: M2} reveals $\Omega_\text{DE} = -\Sigma$. Given this, it's challenging to attribute a physical meaning to the anisotropy, which should behave as a pressure-less fluid. Due to the complexities associated with interpreting this point, we also dismiss it as a viable cosmological epoch in the Universe's expansion history. Thus, we anticipate physical trajectories to begin around the radiation-dominated point (R1) and naturally progress to the matter-dominated point (M1).

The absence of a viable scaling fixed point in our model represents a notable divergence from other models. For instance, in the well-known coupled quintessence model introduced by Amendola in Ref.~\cite{Amendola:1999er, Amendola:1999dr}, the scaling fixed point significantly influences the cosmological dynamics. This impact arises because the coupling modifies the strength of gravity at late times, as reflected in the effective gravitational constant $G_\text{eff}$. However, the lack of cosmologically viable scaling fixed points in our model does not negate the influence of the coupling between the dark sectors on the cosmological evolution, as we will see in a subsequent section.

\subsubsection{\textbf{Dark Energy Dominance}}

\begin{itemize}
\item[$\bullet$] \textbf{(DE1) Isotropic dark energy:}
\end{itemize}

This point corresponds to an isotropic de-Sitter point ($w_\text{DE} = - 1$), with the Universe dominated by the potential of the 2-form field. The only non-zero variable is
\begin{equation}
    v = 1.
\end{equation}
Since $x = 0$, the 2-form field completely decays. Consequently, the exponential potential stabilizes at a constant value, $V_0$. The Universe's behavior at this point is indistinguishable from that driven by the cosmological constant.

To examine the stability of this point, we compute the eigenvalues of the Jacobian $\mathbb{J}$:
\begin{equation}
    \nu_{1, \ldots, 6} = \left\{- 4, -3, -3, -3, - \frac{3}{2} \pm \sqrt{\frac{1}{4} + 12 \lambda} \right\}.
\end{equation}
While the existence of this point is independent of the values of the parameters $\lambda$ and $\mu$, its stability depends on $\lambda$. This point acts as an attractor, with all eigenvalues negative, within the interval
\begin{equation}
\label{Eq: DE1 Stability}
    0 \leq \lambda \leq \frac{1}{6} \quad \land \quad \mu \in \mathbb{R}.
\end{equation}
Therefore, within this parameter space, the fate of the Universe mirrors that of the cosmological constant—a perpetual isotropic de-Sitter expansion.

\begin{itemize}
\item[$\bullet$] \textbf{(DE2) Anisotropic DE:}
\end{itemize}

This point represents the only anisotropic solution for dark energy, with the variables taking the form:
\begin{equation*}
    z = \frac{\sqrt{3}}{2} \sqrt{(1 - 6 \lambda)(2 \lambda - 1)}, \quad v = \sqrt{\frac{3}{2} - 3 \lambda},
\end{equation*}
\begin{equation}
\label{Eq: Point DE2}
    x = \sqrt{\frac{1 - 6 \lambda}{4 \lambda - 2}}, \quad \Sigma = \frac{1}{2} - 3 \lambda, \quad \Omega_\text{DE} = 1.
\end{equation}
Astonishingly, this point corresponds to an \emph{anisotropic de-Sitter solution}, $w_\text{DE} = - 1$, with a non-vanishing $\Sigma$. We will give further details about this exotic point at a further section.

This point exists, i.e., all variables are real, whenever 
\begin{equation} 
\label{Eq: DE2 Existence}
    \frac{1}{6} \leq \lambda < \frac{1}{2}.
\end{equation}
A small anisotropy, as observed, requires $\lambda \approx 1/6$. Conversely, $\lambda = 1/2$ yields an Universe with $\Sigma = - 1$, while the variables associated with the 2-form field vanish ($z = v = x = 0$). Therefore, we only focus on solutions with $\lambda \approx 1/6$, as they define cosmologically viable solutions. The eigenvalues of $\mathbb{J}$ are 
\begin{align}
    \nu_{1, \ldots, 4} &= \{- 4, -3, -3, -3\}, \nonumber \\
    \nu_{5, 6} &= - \frac{3}{2} \left( - 1 \pm \sqrt{5 - 40 \lambda + 112 \lambda^2 - 96 \lambda^3} \right).
\end{align}
From the latter expression, we find that this point acts as an attractor within its region of existence [see Eq.~\eqref{Eq: DE2 Existence}]. Therefore, from the stability analysis, both dark energy points, (DE1) and (DE2), are attractors separated by the bifurcation curve $\lambda = 1/6$, regardless of $\mu$. Hence, the attractor nature of the dark energy dominated points remains unaffected by the coupling's presence. 

To finish this section, we conclude that the cosmological trajectory of the model is 
\begin{center}
    (R1) \ $\rightarrow$ \ (M1) \ $\rightarrow$ \ (DE1) / (DE2),
\end{center}
where the selection of (DE1) or (DE2) depends on the parameter $\lambda$ [see the bounds in Eqs.~\eqref{Eq: DE1 Stability} and \eqref{Eq: DE2 Existence}].

\section{Cosmological Evolution}
\label{Sec: Cosmo Evo}

\subsection{Expansion History}
\label{Sec: Expansion History}

\begin{figure}[t!]
\centering
\includegraphics[width=\linewidth]{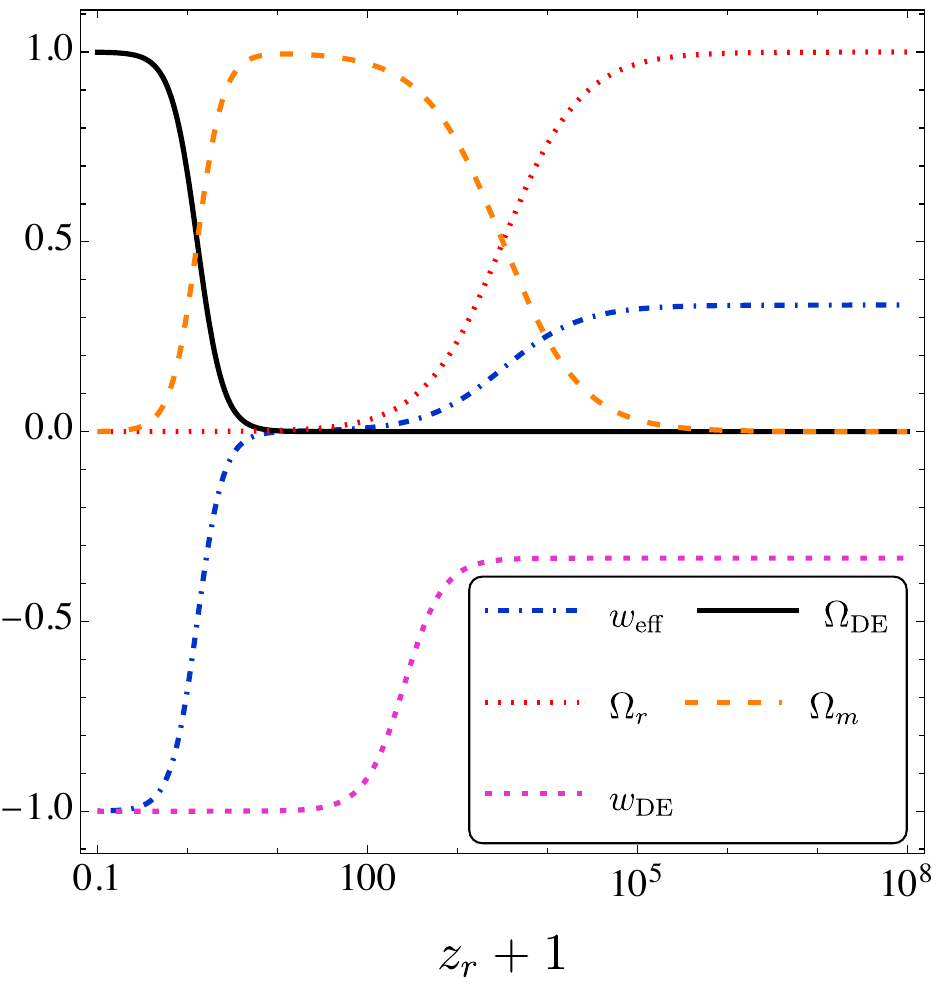}
\caption{(Color online) Evolution of the density parameters, the effective equation of state, and the DE equation of state throughout the entire expansion history of the Universe. The initial conditions, applied in the deep radiation era as detailed in Eq.~\eqref{Eq: Init Cond}, guide the Universe through distinct phases of dominance: early-time radiation dominance (red dotted line), followed by matter dominance (light brown dashed line), and culminating in dark energy dominance (black solid line), which is characterized by a de-Sitter expansion where $w_{\text{eff}} \simeq -1$ (blue dot-dashed line). The equation of state of dark energy (purple small dashed line) exhibits two characteristic behaviors during these phases: $w_{\text{DE}} = -1/3, -1$. }
\label{Fig: Expansion History}
\end{figure}

\begin{figure}[t!]
\centering
\includegraphics[width=\linewidth]{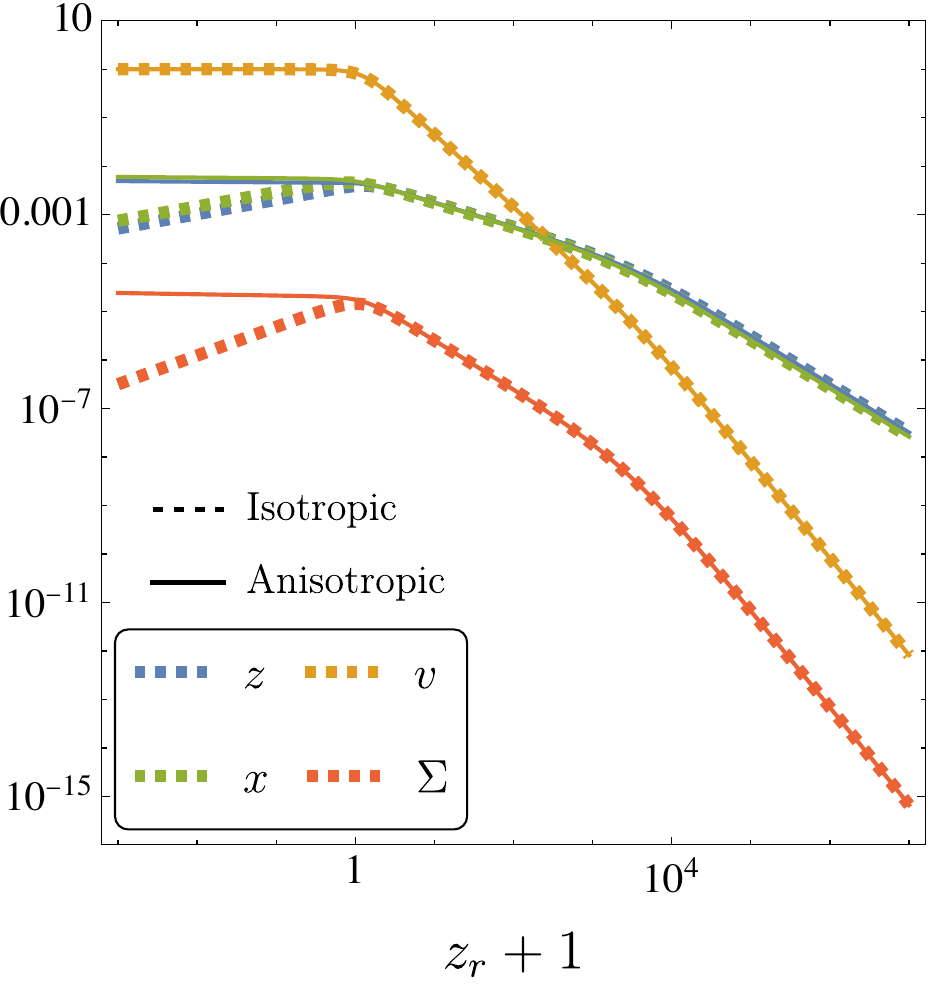}
\caption{(Color online) Cosmological evolution of the variables $z$ (blue), $v$ (yellow), $x$ (green), and $\Sigma$ (orange) under the influence of the two possible attractors. Initially, the evolution of these variables appears indistinguishable at high redshift for both scenarios. However, their paths diverge significantly at later times depending on the selected attractor. In the scenario where the isotropic point (DE1) serves as the attractor (dashed lines), the variable $v$ approaches unity while the others decay. In contrast, when the anisotropic de-Sitter point (DE2) is the attractor (solid lines), all variables tend to stabilize around their respective values at the attractor, as specified in Eq.~\eqref{Eq: Point DE2}.}
\label{Fig: Evo Fields}
\end{figure}

We will proceed to investigate the cosmological dynamics of the fields within the domain where the \emph{anisotropic de-Sitter} solution (DE2) stands as the sole attractor of the system. Therefore, for a small anisotropy at the attractor point and a weak interaction among the dark components, we set the parameters as 
\begin{equation}
\lambda = 0.17, \qquad \mu = 10^{-5}.
\end{equation}

While the parameter values ensure the attractive behavior of (DE2), the expansion history may lead to non-viable cosmologies depending on the chosen initial conditions. For instance, a scenario could arise where the matter-dominated epoch lasts only a few e-folds, insufficient for the structure formation process. Hence, to conduct a successful numerical integration of the autonomous system, we must establish initial conditions guided by physical criteria. To investigate the evolution of shear driven solely by the 2-form field, we assume that an early inflationary phase nullified any anisotropy, rendering the Universe smooth at the onset of the radiation epoch. Deep within the radiation era, such as at $N = -17.28$ corresponding to a redshift $z_r = 3.2 \times 10^7$, we observe $\Omega_{r_i} \sim 1$, with other components being subdominant. Thus, we opt for:
\begin{equation*}
    z_i = x_i = 10^{-8}, \quad v_i = 6 \times 10^{-14}, \quad \Sigma_i = 0,
\end{equation*}
\begin{equation}
\label{Eq: Init Cond}
    \Omega_{b_i} = 10^{-5}, \quad \Omega_{r_i} = 0.9999.
\end{equation}
These initial conditions, set at the specified redshift, are selected to approximate the current state of the Universe, characterized by roughly $\Omega_{r, 0} \approx 10^{-4}$, $\Omega_{m, 0} \approx 0.3$, $\Omega_{\text{DE}, 0} \approx 0.7$, and $w_{\text{DE}, 0} \approx -1$~\cite{Planck:2018vyg}.

In Fig.~\ref{Fig: Expansion History}, we illustrate the dynamical evolution of the density parameters $\Omega_r$, $\Omega_m$, $\Omega_{\text{DE}}$, and the equations of state $w_{\text{eff}}$ and $w_{\text{DE}}$ derived from the numerical solutions of Eqs.~\eqref{Eq: z Eq}-\eqref{Eq: r Eq}. We observe qualitatively that the correct expansion history is reproduced with the initial conditions specified in Eq.\eqref{Eq: Init Cond}. 

A detailed examination of the plot reveals significant insights into the various cosmic epochs. For instance, the radiation-dominated era, characterized by $\Omega_r \approx 1$ and $w_{\text{eff}} \approx 1/3$, spans from $z_r \approx 10^7$ to $z_r \approx 3200$, where the radiation-matter transition occurs~\cite{Planck:2018vyg}. At $z_r = 3200$, we find
\begin{equation*}
    \Omega_r \approx 0.5002, \ \Omega_b + \tilde{\Omega}_c \approx 0.4997, \ \Omega_\text{DE} \approx 5.023 \times 10^{-9},
\end{equation*}
\begin{equation}
    w_\text{eff} \approx 0.1667, \ w_\text{DE} \approx - 0.337,
\end{equation}
and thus we observe that the early contribution of dark energy at this redshift adheres to the big-bang nucleosynthesis (BBN) constraint $\Omega_{\text{DE}} < 0.045$~\cite{Bean:2001wt}. The duration of this radiation phase aligns with constraints provided in Ref.~\cite{Alvarez:2019ues}. Furthermore, we note that the equation of state of dark energy is $w_\text{DE} \approx - 1 / 3$, indicating that this early dark energy fluid is primarily governed by the kinetic term of the 2-form field.

From $z_r \approx 3200$, the Universe transitions to a dust-dominated phase ($\Omega_b + \tilde{\Omega}_c \approx 1$ and $w{\text{eff}} \approx 0$) until $z_r \approx 0.3$, marking the matter-dark energy transition. At $z_r = 50$, we find
\begin{equation*}
\Omega_r \approx 0.0157, \ \Omega_b + \tilde{\Omega}_c \approx 0.984, \ \Omega_\text{DE} \approx 1.62 \times 10^{-5},
\end{equation*}
\begin{equation}
w_\text{eff} \approx 0.0052, \ w_\text{DE} \approx - 0.974,
\end{equation}
highlighting the dominance of the pressure-less matter sector. Notably, the contribution of dark energy at this time adheres to the CMB bound $\Omega_\text{DE} < 0.02$~\cite{Ade:2015rim}.

The dark energy epoch ($\Omega_\text{DE} \approx 1$ and $w_{\text{eff}} < -1/3$) commences around $z_r = 0.3$, consistent with the dynamical system analysis, where (DE2) acts as an attractor. This conclusion is further supported by the agreement between the predicted values of $z$, $x$, $v$, and $\Sigma$ from the dynamical system. In the far future ($z_r \approx -1$),\footnote{Concretely, we computed at $N = 5 \times 10^4$, which is equivalent to \mbox{$z_r \approx -1 + e^{-50000}$}.} our numerical computations yield
\begin{align}
z \approx 0.0994987, &\qquad v \approx 0.994987, \nonumber \\
x \approx 0.123091, &\qquad \Sigma \approx -0.01,
\end{align}
which are consistent with those computed from Eqs.~\eqref{Eq: Point DE2}, reinforcing the validity of our model.

Note that the anisotropic shear plays a relevant role in the dynamics of the model. The effect of the anisotropy can be evidenced by observing the cosmological evolution of the fields. In Fig.~\ref{Fig: Evo Fields}, we plot the time evolution of the variables $z$, $v$, $x$, and $\Sigma$, using the same initial conditions in Eq.~\eqref{Eq: Init Cond} and considering two cases: the isotropic case when $\lambda = 0.1$, agreeing with the attractor condition for (DE1) [Eq.~\eqref{Eq: DE1 Stability}], and the previous studied case when $\lambda = 0.17$ and (DE2) is the only attractor of the system. We note that in both cases, the variables start growing from very small values at high redshift, until they separate at late-times. We see that the isotropic solutions (dashed lines) start decaying for $z_r \approx 0$, excepting $v$ which approximate its attractor value, i.e., $v = 1$. On the other hand, the anisotropic solutions (solid lines), attain to constant values, as predicted from the dynamical analysis, giving rise to an anisotropic de-Sitter phase. 

\begin{figure*}
\centering
\begin{minipage}[b]{.45\textwidth}
\includegraphics[width=\textwidth]{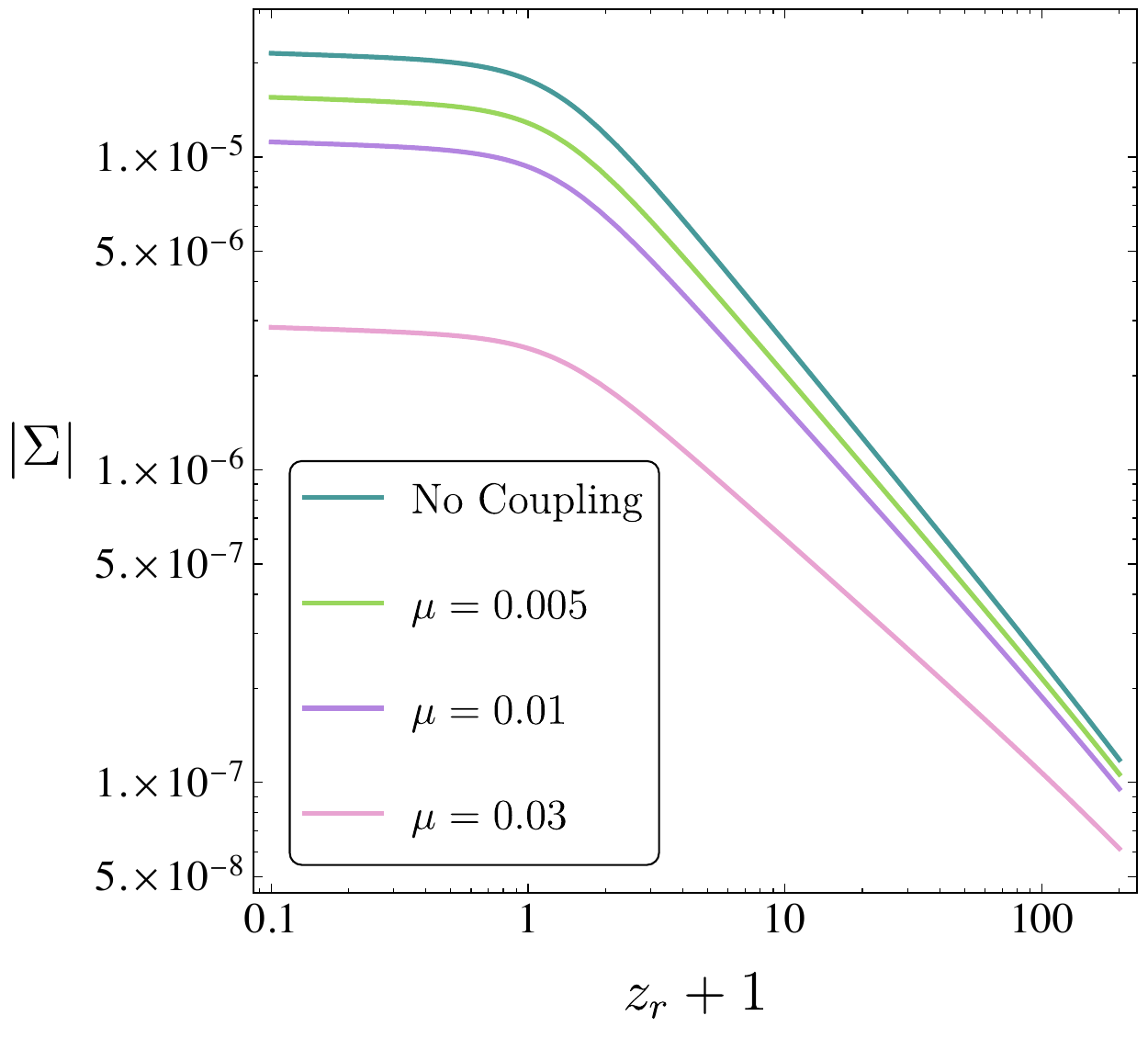}
\end{minipage}
\qquad
\begin{minipage}[b]{.45\textwidth}
\includegraphics[width=\textwidth]{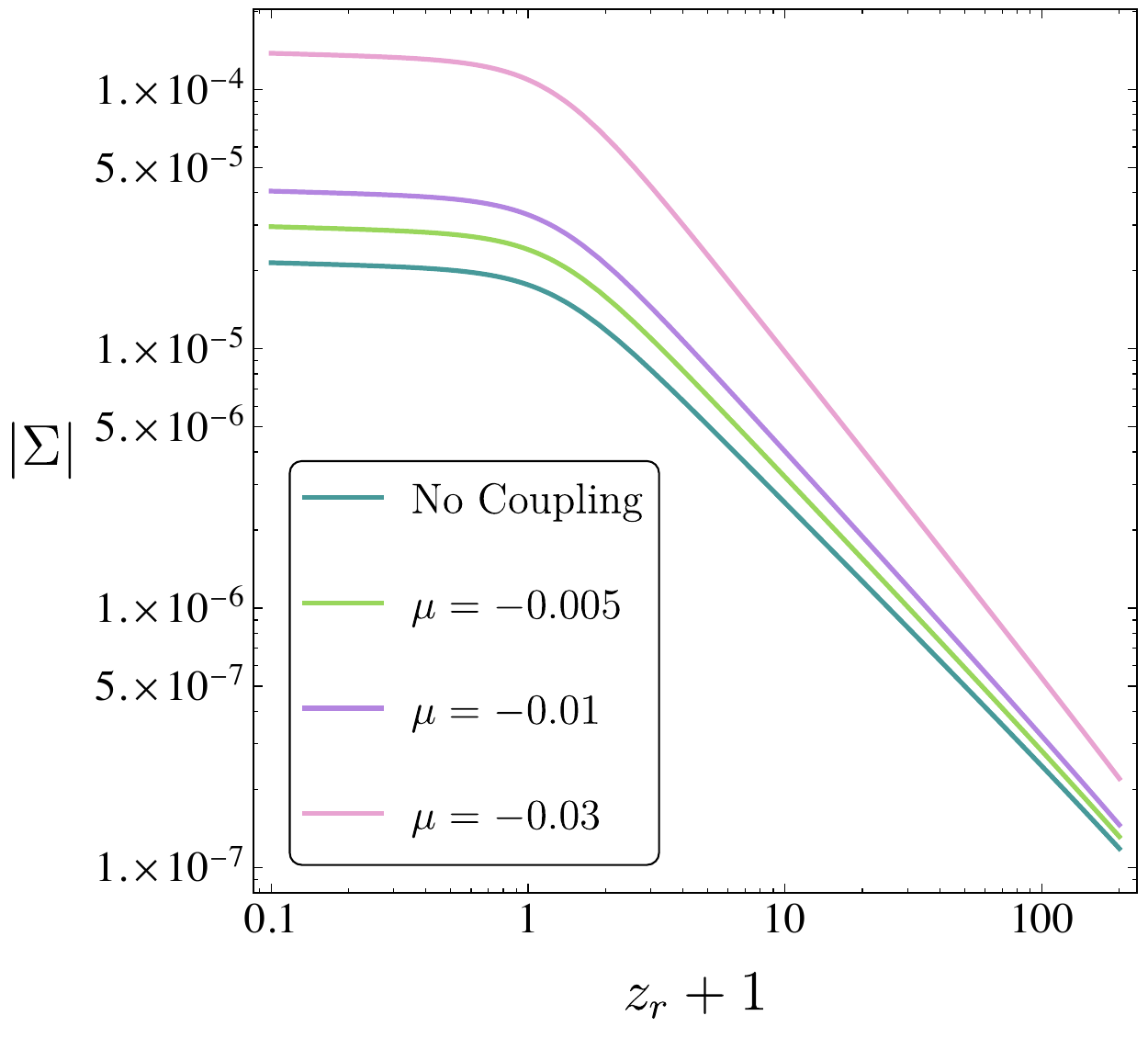}
\end{minipage}
\caption{(Color online) This figure shows how the coupling strength $\mu$ affects the evolution of the shear parameter $\Sigma$, with $\lambda$ set at 0.17 in both scenarios. The sign of $\mu$ significantly influences $\Sigma$'s behavior: (Left) Positive $\mu$ suppresses shear, stabilizing the expansion. (Right) Negative $\mu$ enhances shear, potentially increasing late-time anisotropy.}
\label{Fig: Shear}
\end{figure*}

Our results can be interpreted as follows: the existence of the shear creates a support which hold the evolution of the 2-form field. Let us clarify this point. Note that the evolution equation for the geometrical shear in Eq.~\eqref{Eq: Geometric Shear} in terms of the dynamical variables in Eqs.~\eqref{Eq: Variables} and \eqref{Eq: Variable x} reduces to
\begin{equation}
\label{Eq: Shear in Variables}
    \frac{\ddot{\sigma}}{H^2} + 3 \Sigma = -2z^2 - 4\lambda x^2 v^2 + 4 \mu \tilde{\Omega}_c x^2.
\end{equation}
The latter formula helps us understand one key aspect. Since there is no any CDM-DE scaling attractor, $\tilde{\Omega}_c$ always decays and the shear will tend to negative values, which aligns with the prediction of the dynamical system in Eq.~\eqref{Eq: Point DE2}. When $\Sigma$ remains constant at the attractor (DE2), since $\sigma' = \Sigma$, the geometrical shear is simply expressed as:
\begin{equation}
    \sigma(N) = \int \text{d} N \Sigma \quad \overset{\text{(DE2)}}{\xrightarrow{\hspace{0.6cm}}} \quad - |\Sigma_\text{Att}| N + \sigma_f,
\end{equation}
where $\sigma_f$ is an integration constant, and $\Sigma_\text{Att}$ is the value of the shear at the attractor (DE2) specified in Eq.~\eqref{Eq: Point DE2}. We have verified that this linear relation for $\sigma$ as a function of $N$ holds accurately for $N > 335$, under the given initial conditions. Consequently, the geometrical shear negatively increases linearly with the number of e-folds $N$, supporting the evolution of the 2-form field, since the field $\phi$, its speed $\dot{\phi}$, and the Hubble rate $H$ stabilize at a constant value. This is clear from the DE density, which is given approximately by
\begin{align}
\label{Eq: Density and Anisotropy}
    \rho_\text{DE} &\propto \frac{1}{2}\frac{\dot{\phi}^2}{a^4} a^{4|\Sigma_\text{Att}|} + V_0 e^{-\lambda X/m_\text{P}^2}, \nonumber \\ 
    X &\propto \frac{\phi^2}{a^4} a^{4|\Sigma_\text{Att}|},
\end{align}
where we have considered $\dot{\sigma}$ small, and recalling the relationship between the average scale factor and $N$, i.e., $N \equiv \ln a$. On the other hand, when $\Sigma = 0$, the quadratic term $X$ and the kinetic energy quickly dilutes as $a^{-4}$ and thus the density $\rho_\text{DE}$ sets at a constant value.

The behavior above described is dependent solely on the presence of anisotropy, and the coupling does not alter this conclusion. Our analysis also indicates that the undesirable scaling point (M2), where $\Omega_\text{DE} = - \Sigma$, is not reached during this particular evolutionary trajectory of the Universe, as anticipated. Across a range of values for $\mu$, we consistently observe that the Universe does not stabilize around the scaling point, since $w_\text{DE} \neq 0$ in all scenarios considered. However, as will be discussed next, this does not imply that the coupling lacks an influence on the cosmological evolution of the Universe.

From the evolution equation for the geometrical shear in Eq.~\eqref{Eq: Shear in Variables}, we can extract another important piece of information about our model. Since $\mu$ is unconstrained, it could take on positive or negative values. When $\mu$ is positive, as expected from thermodynamic considerations~\cite{Pavon:2007gt}, the coupling between the 2-form field and CDM suppresses the anisotropy. Conversely, if the coupling strength is negative, the 2-form field is transitioning into CDM and the anisotropy is enhanced. These effects can be corroborated in the left and right panels of Fig.~\ref{Fig: Shear}, respectively, where we have used the same initial conditions as in Eq.~\eqref{Eq: Init Cond}, and fixed $\lambda = 0.17$. Note that a possible enhancement of the anisotropy is constraint by observational bounds on $|\Sigma_0|$, which further precludes the possibility of a strong coupling between the dark sectors. 

\begin{figure}[t!]
\centering
\includegraphics[width=\linewidth]{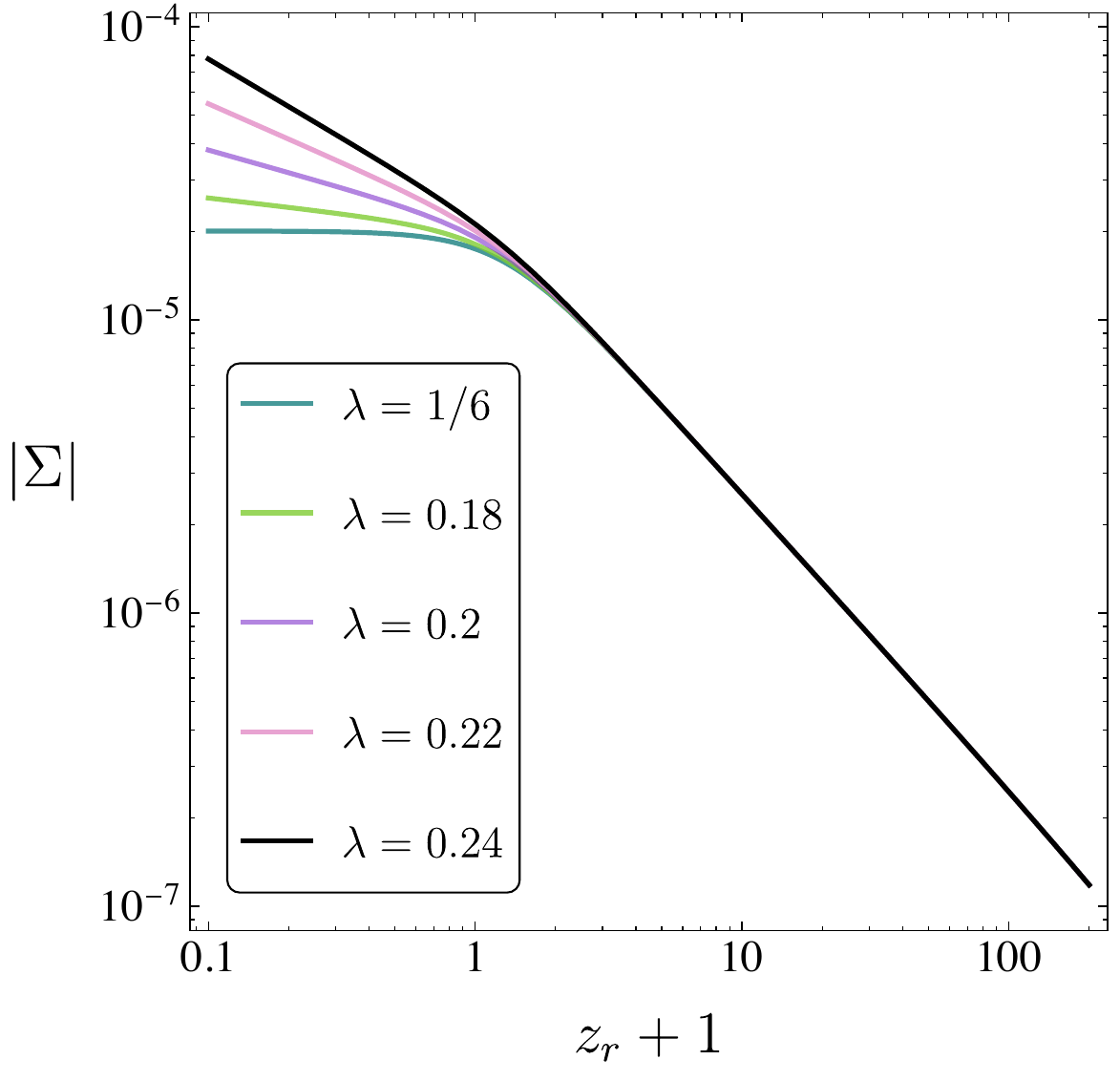}
\caption{(Color online) Influence of the parameter $\lambda$ on the evolution of the shear. The coupling strength is fixed to $\mu = 10^{-5}$. We see that the evolution of $\Sigma$ is independent of $\lambda$ during all the expansion history, except for very-late times when $z_r \sim 0$.}
\label{Fig: Shear lambda}
\end{figure}

In Fig.~\ref{Fig: Shear lambda}, The evolution of the shear is depicted assuming that $\mu = 10^{-5}$ is constant and $\lambda$ varies within the range where (DE2) acts as the attractor. We observe that $|\Sigma|$ increases with $\lambda$ and attains non-negligible values at present. Although $\Sigma$ reaches appreciable levels today, its growth intensifies towards late times (as $z_r \sim 0$), suggesting that this shear may not significantly alter the overall expansion history of the Universe. Contrary to the case when $\mu$ varies (Fig.~\ref{Fig: Shear}). since the different shears diverge from higher redshifts ($z_r < 100$). This configuration shows that the shear begins to increase from around $z_r \sim 100$, indicating that $\Sigma$ could potentially impact the matter-dominated epoch. Such a dynamic could influence the formation of structures, potentially offering a way to mitigate existing tensions related to this process~\cite{Pitrou:2015iya, BeltranAlmeida:2021ywl}. Nonetheless, for both cases, the predicted values for the shear at present align well with observational constraints, which cap $|\Sigma_0|$ at $\mathcal{O}(0.001)$~\cite{Campanelli:2010zx, Amirhashchi:2018nxl}. 

We want to emphasize that the presence of a non-vanishing shear is a robust prediction of our model. This occurs even if the initial shear $\Sigma_i = 0$, suggesting that the final state of the Universe could involve a late-time anisotropic de-Sitter expansion. Consequently, we confirm that the cosmological trajectory of the model follows the sequence:
\begin{center}
(R1) \ $\rightarrow$ \ (M1) \ $\rightarrow$ \ (DE1) / (DE2),
\end{center}
where the transition into either (DE1) or (DE2) is contingent upon the parameter $\lambda$, as specified within the constraints set by Eqs.~\eqref{Eq: DE1 Stability} and \eqref{Eq: DE2 Existence}. 

\subsection{Impact on the CMB Quadrupole}

As mentioned in the introduction, the presence of an anisotropic dark energy component can influence the CMB temperature anisotropies due to the anisotropic nature of the redshift at decoupling. This means that different expansion rates occur around the last scattering surface. In the expansion of the CMB temperature fluctuations into multipoles, the primary correction is a quadrupole term~\cite{Koivisto:2008ig, Thorsrud:2012mu}. Higher multipoles are significantly suppressed and effectively negligible as long as the shear remains small. The quadrupole contribution can be quantified as~\cite{Appleby:2009za}:
\begin{equation}
    \frac{\delta T}{T} (\boldsymbol{\hat{n}}) \Bigg|_\text{quad} = - \int_{t_\text{dec}}^{t_0} \text{d} t \, \sigma_{i j} \hat{n}^i \hat{n}^j.
\end{equation}
Here, $t_\text{dec}$ denotes the cosmic time at decoupling ($z_r \approx 1090$), $t_0$ represents today's time ($z_r = 0$), $\hat{n}^i$ is the vector pointing in the direction of the line-of-sight, and $\sigma_{i j}$ is the geometric shear tensor defined as:
\begin{equation}
    \sigma_{i j} \equiv \dot{\gamma}_{ij} / 2, \quad \gamma_{ij} \equiv e^{2\beta_i} \delta_{i j}.
\end{equation}
In the LRS configuration, we have $\beta_1 = -2\sigma$ and $\beta_2 = \beta_3 = \sigma$. Thus, the product $\sigma_{i j} \hat{n}^i \hat{n}^j$ is at most of order $\dot{\sigma}$, whenever the shear remains small. The contribution from the shear quadrupole is constrained by:
\begin{equation}
\left| \frac{\delta T}{T} (\boldsymbol{\hat{n}})\right|_\text{quad} \lesssim | \sigma_\text{dec} - \sigma_0 |,
\end{equation}
where $\sigma_\text{dec}$ is the geometrical shear at decoupling, and $\sigma_0$ is its current value. Since $\Sigma = \sigma'$, numerical integration of $\Sigma$ from the solutions of the autonomous system allows us to compute $\sigma$. Using the same initial conditions in Eq.~\eqref{Eq: Init Cond} but with parameters set for different attractors, we find for (DE2) with $\lambda = 0.17$ and $\mu = 10^{-5}$:
\begin{equation}
| \sigma_\text{dec} - \sigma_0 | \approx 2.27 \times 10^{-5},
\end{equation}
and for (DE1) with $\lambda = 0.01$ and $\mu = 10^{-5}$:
\begin{equation}
| \sigma_\text{dec} - \sigma_0 | \approx 2.16 \times 10^{-5}.
\end{equation}
In Ref.~\cite{Appleby:2009za}, it has been claimed that the CMB quadrupole anomaly could be explained by an anisotropic dark energy whenever the contribution of the geometrical shear is constrained by the conservative bound $| \sigma_\text{dec} - \sigma_0 | < 10^{-4}$. We see that our estimation perfectly aligns with this bound, even when the isotropic point (DE1) is chosen as the attractor. Notably, this result underlines that although the eternal accelerated expansion of the Universe might be isotropic, the present-day anisotropy could still be significant, as emphasized in Ref.~\cite{Orjuela-Quintana:2020klr}, leading to observable signatures in the CMB quadrupole.

\subsection{On the Existence of Anisotropic de-Sitter Solutions}

Under very general and physically motivated conditions, any expanding Universe will evolve toward a locally de-Sitter Universe, according to the famous cosmic no-hair conjecture~\cite{Hawking:1981fz}. Wald proved this conjecture for all Bianchi models, assuming a positive cosmological constant and that all matter fluids obey the strong energy condition (SEC) and the dominant energy condition (DEC)~\cite{Wald:1983ky}. These conditions, when expressed in terms of the density $\rho$ and pressure $p$ of a perfect fluid, are given by:
\begin{equation}
\text{SEC: } \quad \rho + 3p \geq 0, \qquad \text{DEC: } \quad \rho \geq |p|.
\end{equation}
Thus, it is expected that fluids violating any of these conditions might also evade the no-hair conjecture, potentially supporting anisotropy. In light of this, several models allowing for prolonged anisotropic expansion phases have been proposed~\cite{Barrow:1987ia, Maleknejad:2012as}.

Specifically related to our work, Kaloper studied the cosmic no-hair conjecture under the assumption that the matter sector includes the kinetic term associated with the Kalb-Ramond 2-form field, alongside a positive cosmological constant. The action of this model is given by~\cite{Kaloper:1991rw}:
\begin{equation}
\label{Eq: Kaloper Action}
S = \int \text{d}^4 x \sqrt{-g} \left[ \frac{m_\text{P}^2}{2}R - \frac{1}{12} H_{\mu\nu\rho} H^{\mu\nu\rho} - \Lambda \right].
\end{equation}
Kaloper demonstrated a breakdown of the cosmic no-hair conjecture for specific Bianchi cosmologies, such as the diagonal Bianchi type-II geometry. Notably, Bianchi I and V Universes do not exhibit such a breakdown, meaning the conjecture holds in these spacetimes.

Our model, expressed in Eq.~\eqref{Eq: Action}, resembles Kaloper's action in Eq.~\eqref{Eq: Kaloper Action}, albeit with the cosmological constant replaced by a potential that encodes auto-interactions of the 2-form field. However, our model leads to different conclusions compared to those of Kaloper.

In the isotropic attractor (DE1), we observe that the field decays and its potential stabilizes at a constant value $V_0$, effectively acting as a cosmological constant. This aligns with Kaloper's results for a Bianchi I spacetime. However, the potential also presents another possibility in the anisotropic attractor (DE2). As detailed in Sec.~\ref{Sec: Expansion History}, at this point, the field does not decay, and the anisotropy remains constant. This represents a clear violation of Wald's theorem in a Bianchi I spacetime, suggesting that a 2-form field could indeed source an anisotropic accelerated expansion of the Universe across various anisotropic cosmologies.

While this work focuses on the simplest realization of the Bianchi geometries, specifically the LRS Bianchi I spacetime, most general 2-form field profiles have been explored in the literature~\cite{Normann:2017aav, BeltranAlmeida:2018nin}. However, the impact of an auto-interacting potential has not yet been considered. We aim to address this gap in future work.

We have discovered that both attractors in our model lead to a de-Sitter expansion phase, wherein the average Hubble rate $H$ remains constant, but there is a non-zero shear component. This introduces an observational degeneracy between the $\Lambda$CDM model and our model, as the observation of $w_\text{DE, 0} = -1$ does not necessarily imply $|\Sigma_0| = 0$. Consequently, it becomes challenging for observations to distinguish between isotropic and anisotropic late-time accelerated expansions based solely on this parameter. This degeneracy is akin to that found in designer models of dark energy, where the dark energy density and pressure are tailored to mimic the Hubble parameter evolution of the $\Lambda$CDM model. As a result, any observable dependent on $H(z)$, such as the luminosity distance, will not be able to differentiate these models~\cite{Sahni:2002fz, Zimdahl:2003wg}.

It is important to emphasize that the existence of the \emph{anisotropic de-Sitter} attractor is contingent upon the framework used to describe the dark energy stress tensor~\cite{Akarsu:2010zm}. Generally, we employ the standard decomposition of the ``$ij$'' components of the energy tensor in terms of an ``isotropic pressure'' and a traceless shear tensor, expressed as:
\begin{equation}
    p_\text{DE} = \frac{1}{3} T^{i (B)}_{\ i}, \quad \Pi^i_{\ j} \equiv T^{i (B)}_{\ j} - p_\text{DE} \delta^i_{\ j}.
\end{equation}
But, an alternative decomposition is also feasible:
\begin{equation}
    T^{i (B)}_{\ j} \equiv \text{diag}(-\rho_\text{DE}, p_\parallel, p_\perp, p_\perp),
\end{equation}
where different pressures are considered. In our scenario, the expressions for $p_\parallel$ and $p_\perp$ would be derived from $T^{1 (B)}_{\ 1}$ and $T^{2 (B)}_{\ 2}$, respectively, as shown in Eq.~\eqref{Eq: T11} and Eq.~\eqref{Eq: T22}. Utilizing this scheme, the dark energy fluid is characterized by two distinct equations of state parameters, $w_\parallel$ and $w_\perp$, which generally do not align with de-Sitter solutions.

Using the ``\emph{parallel-perpendicular}'' decomposition, we can rewrite the evolution equation for the shear in Eq.~\eqref{Eq: Geometric Shear} as:
\begin{equation}
    3m_\text{P}^2\ddot{\sigma} + 9 m_\text{P}^2 H \dot{\sigma} = p_\perp - p_\parallel,
\end{equation}
and thus the behavior of the shear is controlled by the difference in the pressures. However, note that the derivation of this equation and its further analysis, as we presented in previous sections, do not actually require the definition of $p_\parallel$ and $p_\perp$. Although both formulations are equivalent, we believe that the standard decomposition considering a shear tensor and an average pressure is easier to interpret. For instance, the standard decomposition is generally used in the standard perturbation theory scheme.

\section{Difference with respect to a Vector Field}
\label{Sec: Difference wrt a Vector Field}

In the $\Lambda$CDM scenario, Wald's theorem suggests that any expanding Universe will eventually evolve toward a de-Sitter phase characterized by constant expansion rate. In the simplest case of the LRS Bianchi-I spacetime, this can be exemplified since the evolution of the geometrical shear is given by:
\begin{equation}
    \ddot{\sigma} + 3 H \dot{\sigma} = 0.
\end{equation}
indicating that the cosmological constant is not able to support the shear, and thus it decays quickly as $a^{-3}$. 

The universal fate toward isotropy can be altered if an anisotropy-sourcing field is present, effectively modifying the above equation by adding a source term to the right-hand side. Typical mechanisms that support anisotropic accelerated expansion involve a scalar field driving expansion coupled with an anisotropy-supporting field, such as vector fields or 2-form fields. However, alternative configurations such as non-homogeneous scalar fields or non-abelian gauge fields have also been shown to sustain anisotropic expansion on their own.

Our investigation reveals that an anisotropic accelerated expansion can be uniquely sustained by a 2-form field's dynamics with auto-interactions defined by its potential $V$. Previous studies have demonstrated that a 1-form field, or vector field, can support isotropic accelerated expansion~\cite{Landim:2016dxh, Gomez:2020sfz}. This raises a critical question: Can a vector field also support anisotropic accelerated expansion?

To explore this, consider the action for a vector field:
\begin{equation}
    S_\text{v} = \int \text{d}^4 x \sqrt{-g} \left[ \frac{m_\text{P}^2}{2} R - \frac{1}{4} F_{\mu\nu} F^{\mu\nu} - V_A (X_A) \right],
\end{equation}
where $F_{\mu\nu} \equiv \nabla_\mu A_\nu - \nabla_\nu A_\mu$ represents the field strength tensor for the vector field $A_\mu$, and $V_A$ is a potential typically chosen to have an exponential form:
\begin{equation}
    V_A (X_A) \equiv V_0 e^{-\lambda_A X_A/m_\text{P}^2}, \quad X_A \equiv - \frac{1}{2} A_\mu A^\mu,
\end{equation}
with $\lambda_A$ as a positive constant. Note that this action resembles the action of a 2-form field as in Eq.~\eqref{Eq: Action}.

Assuming a vector field profile compatible with the LRS Bianchi I metric: 
\begin{equation}
    A_\mu \equiv (0, \psi(t), 0, 0),
\end{equation}
where $\psi(t)$ is a time-dependent scalar field, leads to the following evolution equation for the shear:
\begin{equation}
    \ddot{\sigma} + 3 H \dot{\sigma} = \frac{\dot{\psi}^2 e^{4\sigma}}{3 m_\text{P}^2 a^2} - \lambda_A \left(\frac{\psi^2 e^{4\sigma}}{m_\text{P}^2 a^2}\right) \frac{V_A}{3 m_\text{P}^2}.
\end{equation}
This equation is very similar to the equation presented in Eq.~\eqref{Eq: Geometric Shear} for the 2-form field. Note also that if the vector field sources the accelerated expansion, its potential energy must dominate and thus the first term on the right-hand side of the latter equation should be negligible. Therefore, the shear tends to negative values, as in the case of the 2-form field. However, this also implies that the vector field will not be able to sustain prolonged phases of anisotropic expansion. Neglecting second order terms in $\dot{\sigma}$, the density associated to the vector field is given by
\begin{equation}
    \rho_A \approx \frac{\dot{\psi}^2 e^{4\sigma}}{2a^2} + V(X_A), \quad X_A = -\frac{1}{2} \frac{\psi^2 e^{4\sigma}}{a^2}.
\end{equation}
Since $\sigma(t)$ tends to negative values, this density will exponentially reach the constant value of its potential. Contrary to the 2-form field, as shown in a previous section, its density is fed by the negative anisotropy [see Eqs.~\eqref{Eq: Density and Anisotropy}].

Therefore, the 1-form model should only possess one attractor point at which the vector field completely decays, its potential is flattened, and in general becomes indistinguishable from the $\Lambda$CDM scenario. A rigorous confirmation of this claim, however, is beyond the scope of the present work and will be addressed in a future project.

\section{Cosmological Constraints}
\label{Sec: Cosmological Constraints}

\subsection{Observational Data}

We constrain the parameter space of our model by fitting it to recent observational data. Specifically, we employ: \\ \\
$\bullet$ \textbf{SnIa:} Type Ia supernovae (SnIa) provide measurements of the distance modulus $\mu_\text{Sn}$, defined as:
\begin{equation}
    \mu_\text{Sn} \equiv m_B - M_B = 5 \log_{10} \left[ \frac{d_L}{\text{Mpc}} \right] + 25,
\end{equation}
where $m_B$ is the apparent magnitude and $M_B$ the absolute magnitude. The luminosity distance $d_L$ as a function of the redshift $z_r$ is computed through:
\begin{equation}
    d_L (z_r) = (1 + z_r) \int_0^{z_r} \frac{\text{d}z'}{H(z')}.
\end{equation}
We adopt the Hubble rate $H(z)$ from Eq.~\eqref{Eq: Friedman 1 Eq} and utilize the Pantheon+ dataset, which includes 1701 light curves from 1550 distinct SnIa across redshifts $z_r \in [0.001, 2.26]$. This dataset incorporates the Cepheid host calibrator of distance and covariance from SH0ES~\cite{Brout:2022vxf, Riess:2021jrx}. \\ \\
$\bullet$ \textbf{BAO:} Baryon acoustic oscillations (BAO) provide constraints on the ratio $D_V/r_d$, where
\begin{equation}
    D_V (z_r) = \left[ c z_r (1 + z_r)^2 \frac{D_A^2(z_r)}{H(z_r)} \right]^{1/3},
\end{equation}
represents the spherically averaged distance calculated from the diameter angular distance $D_A(z)$, and
\begin{equation}
    r_s(z_\text{d}) \equiv r_\text{d} = \int_{z_\text{d}}^\infty \text{d} z_r \frac{c_s(z_r)}{H(z_r)},
\end{equation}
is the sound horizon $r_s$ at the drag epoch, which occurs at redshift $z_\text{d}$, with $c_s$ being the sound speed in the photon-baryon fluid. Here, we employ observational data from 6dFGS~\cite{Beutler:2011} and SDSS DR7 MGS~\cite{Ross:2014qpa}. \\ \\
$\bullet$ \textbf{H(z):} Direct measurements of the Universe's expansion rate can be made by observing the BAO peak in the radial direction from galaxy or quasar clustering, or via the differential age method, analyzing the redshift drift of distant objects over significant periods, as the Hubble rate is given by:
\begin{equation}
    H(z_r) = -(1 + z_r)\frac{\text{d} z_r}{\text{d} t}.
\end{equation}
For this analysis, we utilize data compiled in Ref.~\cite{Guo:2015gpa} and from Ref.~\cite{Moresco:2016mzx}. \\ \\
$\bullet$ \textbf{CMB:} The presence of dark energy modifies certain properties of the CMB peaks. These modifications are captured by the shift parameters~\cite{Wang:2013mha}:
\begin{align}
    R &\equiv \sqrt{\Omega_{m0} H_0^2} r(z_\star)/c, \\
    \ell_a &\equiv \pi r(z_\star)/r_\text{s}(z_\star),
\end{align}
where $r(z_\star)$ is the comoving distance to the photon decoupling surface at redshift $z_\star$, $\Omega_{m0}$ and $H_0$ represent the current values of the matter density parameter and the Hubble rate, respectively. Shift parameter values are taken from Planck data~\cite{Planck:2018vyg}, incorporating the reduced baryon density parameter $\omega_b$, and the effective number of relativistic species $N_\text{eff}$. 

As discussed in the preceding sections, the influence of the shear on the cosmological dynamics becomes significant only at very late times ($z \lesssim 10$), as this term is sourced by dark energy, which has only recently become a dominant component of the cosmic energy budget. Moreover, the effect of the shear can be characterized as a ``correction term'' to the Hubble parameter [see Eq.~\eqref{Eq: Friedman 1 Eq}], given that $\Sigma$ remains small throughout the entire history of the universe. This justifies the use of these datasets, despite their initial adaptation within a FLRW framework.

\subsection{Constraints on the Parameter Space}

\begin{center}
\begin{table*}
\begin{centering}
\begin{tblr}{X[2, c] X[1.2, c] X[1.2, c] X[1.2, c] X[1.2,c] X[1.5, c]}
\hline \hline
CMB & $h$ & $N_\text{eff}$ & $100\omega_b$ & $\Omega_c$ & $10^{4}\Sigma_0$ \\
\hline \hline
$\Lambda$CDM & $0.655 \pm 0.019$ & $2.92 \pm 0.20$ & $2.22 \pm 0.02$ & $0.275 \pm 0.016$ & $-$ \\
Coupled 2-Form & $0.654 \pm 0.019$ & $2.92 \pm 0.19$ & $2.22 \pm 0.02$ & $0.275 \pm 0.015$ & $-3.75^{+3.73}_{-0.52}$ \\
\hline \hline
CMB+SnIa+BAO+Hz & $h$ & $N_\text{eff}$ & $100\omega_b$ & $\Omega_c$ & $10^{4}\Sigma_0$ \\
\hline \hline
$\Lambda$CDM & $0.705 \pm 0.007$ & $3.37 \pm 0.12$ & $2.28 \pm 0.01$ & $0.244 \pm 0.007$ & $-$ \\
Coupled 2-Form & $0.705 \pm 0.007$ & $3.37 \pm 0.12$ & $2.28 \pm 0.01$ & $0.244 \pm 0.007$ & $-3.50^{+3.48}_{-0.53}$ \\
\hline \hline
\end{tblr}
\par\end{centering}
\caption{The best-fit parameters for $\Lambda$CDM and our model from two analysis: CMB data only, and a joint dataset comprising CMB, SnIa, BAO, and H(z) observations. Although the parameters of our model, $\lambda$ and $\mu$, are not constrained, remarkably, the derived today's anisotropy is tightly constrained to be approximately $\Sigma_0 \approx - 10^{-4}$.}
\label{Tab: Models}
\end{table*}
\par\end{center}

\begin{figure*}
\centering
\begin{minipage}[b]{0.85\textwidth}
\includegraphics[width=\textwidth]{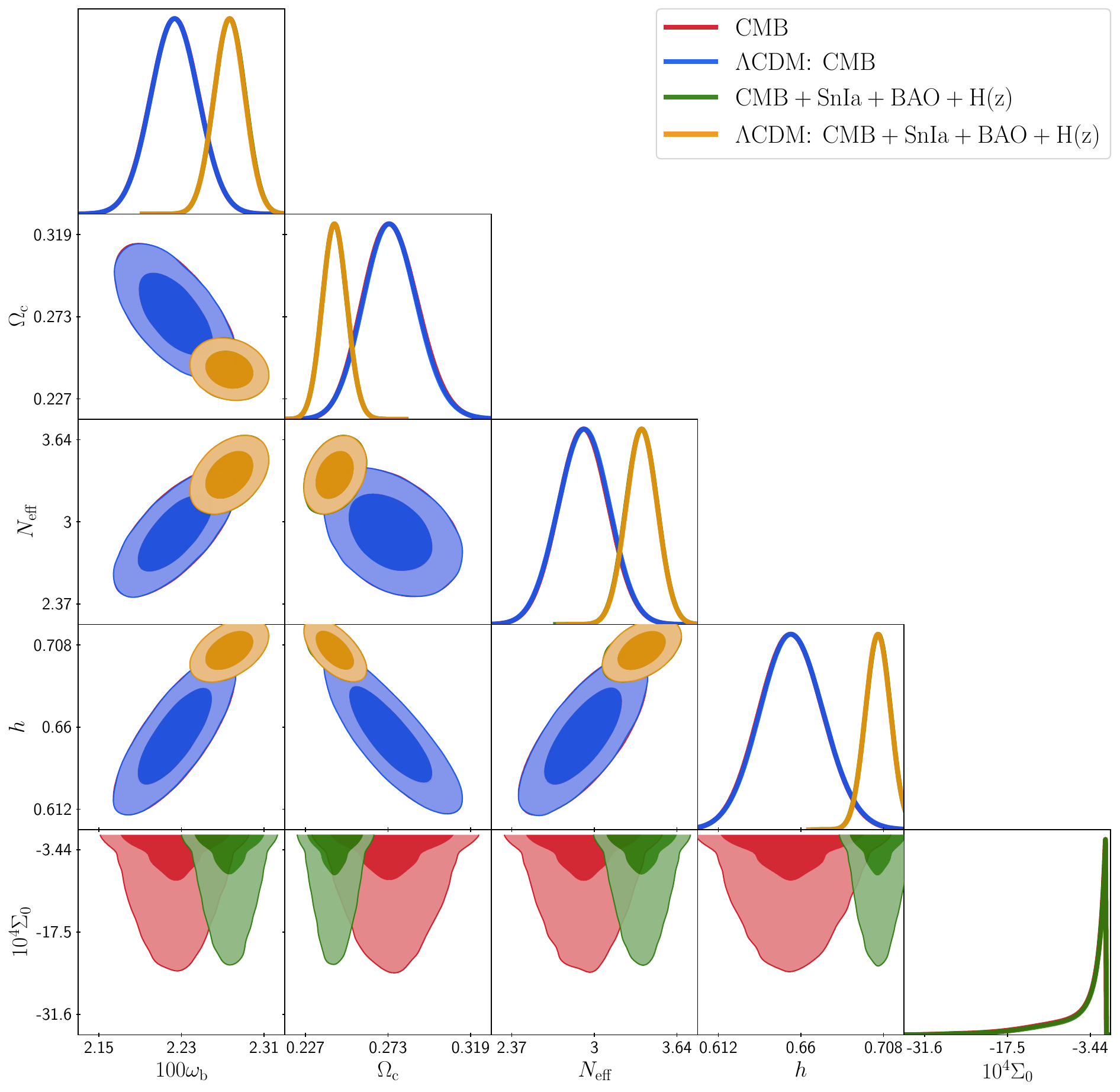}
\end{minipage}
\caption{(Color online) The 68.3\% and 95.4\% confidence contours and the 1D marginalized likelihoods for the free parameters $\{\omega_b, \Omega_c, h, N_\text{eff}\}$ of the $\Lambda$CDM model and our anisotropic and interacting scenario, using background observations. The current shear is presented as a derived parameter. It is clear that our theoretical proposal is qualitatively degenerate with respect to $\Lambda$CDM.}
\label{Fig: Contours}
\end{figure*}

To constrain the parameter space of our model using the above-described observational data, we compute the Hubble rate $H(z_r)$ as per Eq.~\eqref{Eq: Friedman 1 Eq}. Since this function is not directly retrieved from the solution of the dynamical system, we solve the coupled set of differential equations given by the Friedman equations [Eqs.~\eqref{Eq: Friedman 1 Eq} and \eqref{Eq: Friedman 2 Eq}], the evolution equation for the shear [Eq.~\eqref{Eq: Geometric Shear}], and the equation of motion for the 2-form field [Eq.~\eqref{Eq: 2-form EoM}]. In order to do so, we introduce this system of differential equations into the well-known Boltzmann solver \texttt{CLASS}~\cite{Blass:2011}.

After incorporating our model into \texttt{CLASS}, we proceed to explore the parameter space using the Metropolis-Hastings algorithm, a widely employed Markov Chain Monte Carlo (MCMC) method, facilitated by the \texttt{Monte Python} code~\cite{Audren:2013, Brinckmann:2018cvx}. The resulting chains generated from the MCMC runs encapsulate the probability distribution of the model parameters, which are subsequently scrutinized and interpreted using the analytical capabilities of the \texttt{Monte Python} framework. This algorithmic approach allows for efficient sampling of the multidimensional parameter space, which is encoded in the parameter vector of our model: 
\begin{equation}
    \mathcal{P} \equiv \{\omega_b, \Omega_c, h, N_\text{eff}, \lambda, \mu \},
\end{equation}
where $h$ is the reduced Hubble parameter defined trough $H_0 = 100 h \, \text{km} \, s^{-1} \, \text{Mpc}^{-1}$. The cosmological parameters are constrained within these flat priors: $100\omega_b \in [2.0, 2.5]$, $\Omega_c \in [0.1, 1.0]$, $h \in [0.1, 1.0]$, $N_\text{eff} \in [1.0, 5.0]$, $\lambda \in [0, 0.5]$, and $\mu \in [-0.05, 0.05]$. We also introduce the present-day value of the anisotropy $\Sigma_0$ as a derived parameter.

\begin{figure}[t!]
\centering
\includegraphics[width=\linewidth]{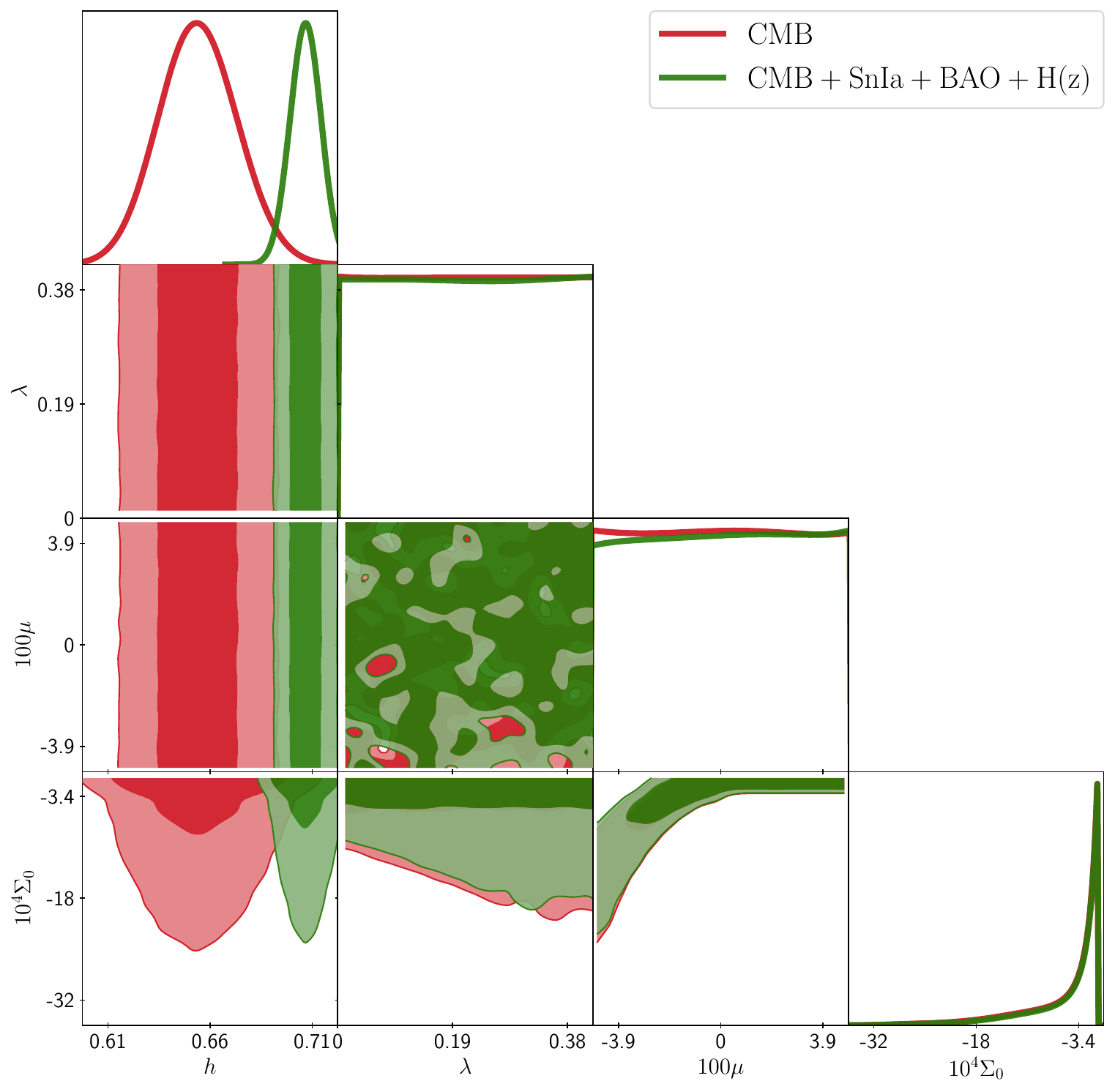}
\caption{The 68.3\% and 95.4\% confidence contours and the 1D marginalized likelihoods for the free parameters of the anisotropic scenario $\{\lambda, \mu\}$ using background observations. The current shear is presented as a derived parameter. It is clear that the parameters $\lambda$ and $\mu$ are unconstrained by the employed datasets, although $\Sigma_0$ can be determined.}
\label{Fig: Unconstrained Params}
\end{figure}

For cosmological inference regarding best-fit parameters, we consider two observational datasets, namely, CMB data alone, and the joint dataset comprising CMB, SnIa, BAO, and H(z) observations, for two theoretical cases: the standard $\Lambda$CDM model and our interacting scenario. Figure~\ref{Fig: Contours} displays the 2D contours for the 68.3\% and 95.4\% confidence levels, alongside the corresponding 1D marginalized likelihoods for each parameter, excluding $\lambda$ and $\mu$ due to dataset insensitivity. The best-fit parameters are detailed in Table~\ref{Tab: Models}. Note that $h \approx 0.7$ when all datasets are considered. Although this represents an alleviation to the $H_0$ tension, this amelioration is artificial since the dataset is dominated by the SnIa data. Therefore, a more careful analysis is needed in order to draw a precise conclusion on this subject. On the other hand, we want to highlight that, although $\lambda$ and $\mu$ are unbounded by these datasets (see Fig.~\ref{Fig: Unconstrained Params}), the constraints placed on the late-time anisotropy are remarkably precise, as demonstrated by the following tightly bound estimates:
\begin{align}
    \text{CMB: } &\quad \Sigma_0 = - 3.75^{+3.73}_{-0.52} \times 10^{-4}, \nonumber \\
    \text{Full data: } &\quad \Sigma_0 = - 3.50^{+3.48}_{-0.53} \times 10^{-4}. \nonumber
\end{align}

To evaluate the statistical significance of our constraints on the cosmological model parameters, we employ the Akaike information criterion (AIC). This criterion is a widely used statistical measure that quantifies the goodness of fit of a model while penalizing excessive complexity to avoid overfitting~\cite{Akaike:1974}. The AIC is defined as:
\begin{equation}
\text{AIC} = - 2\ln \mathcal{L}_\text{max} + 2k_p + \frac{2k_p (k_p + 1)}{N_\text{data} - k_p - 1},
\end{equation}
where $\mathcal{L}_\text{max}$ is the maximum likelihood of the model, $k_p$ the number of free parameters, and $N_\text{data}$ the number of data points. In few words, the lower the AIC value, the better the balance between goodness of fit and model simplicity. The dataset distribution includes 4 data points from CMB, 1701 from SnIa, 2 from BAO, and 36 from $H(z)$, giving a total of $N_{\text{data}} = 1743$ data points. The maximum likelihood estimations are derived using the \texttt{Monte Python} software. In terms of model parameters, the standard $\Lambda$CDM model has 4 free parameters: $\{\omega_b, \Omega_c, h, N_{\text{eff}}\}$. Our model includes two additional parameters, $\lambda$ and $\mu$, resulting in a total of 6 parameters. This inclusion allows us to test whether the additional complexity introduced by $\lambda$ and $\mu$ is justified by a sufficiently improved fit to the data as measured by the AIC.

\begin{table}[t!]
\setlength{\tabcolsep}{12pt}
\begin{tabular}{ccccc}
\hline \hline
Model & $-\ln \mathcal{L}_\text{max}$ & $k_p$ &AIC & $\Delta$ AIC \\
\hline
\hline
$\Lambda$CDM & 670.37 & 4 & 1348.76 & 0.0 \\
\hline
2-Form & 670.38 & 6 & 1352.81 & 4.04 \\
\hline
\end{tabular}
\caption{AIC parameters for the $\Lambda$CDM model and our interacting 2-for field-CDM model. Since $\Delta\text{AIC} \sim 4$ for our model, it is disfavored with respect to $\Lambda$CDM.}
\label{Table: AIC}
\end{table}

To effectively apply the AIC for model selection, we calculate the pairwise differences between models using the formula \mbox{$\Delta \text{AIC} = \text{AIC}_{\text{model}} - \text{AIC}_{\text{min}}$}, where $\text{AIC}_{\text{min}}$ is the AIC value for the best-fitting model among those being compared. This difference, $\Delta \text{AIC}$, helps to quantify the relative quality of each model under consideration.

Interpretation of $\Delta \text{AIC}$ values typically follows Jeffreys' scale, which categorizes the strength of evidence against models with higher AIC values. According to this scale:
\begin{itemize}
    \item A $\Delta \text{AIC}$ value between 4 and 7 suggests positive evidence against the model with the higher $\text{AIC}_{\text{model}}$,
    \item A $\Delta \text{AIC}$ of 10 or more indicates strong evidence against the higher AIC model,
    \item A $\Delta \text{AIC}$ of 2 or less implies that the models have statistically equivalent support.
\end{itemize}
However, as noted in Reference~\cite{Nesseris:2012cq}, the application of Jeffreys' scale can sometimes lead to misleading conclusions. Therefore, while $\Delta \text{AIC}$ provides a useful heuristic for comparing models, it should always be interpreted with caution, considering the specific context and assumptions of the analysis. In Table~\ref{Table: AIC}, we present the computed values of the AIC for both the standard $\Lambda$CDM model and our proposed model. Our analysis reveals that both models achieve a comparable level of maximization of the likelihood. However, the inclusion of the extra parameters $\lambda$ and $\mu$ in our model introduces a complexity penalty as quantified by the AIC. The statistical difference between the two models is approximately 4, indicating a moderate preference for $\Lambda$CDM. This result suggests that the additional complexity introduced by the extra parameters in our model does not sufficiently improve the fit to the observational data to justify their inclusion over the simpler $\Lambda$CDM model.

\section{Conclusions} 
\label{Sec: Conclusions}

The observed discrepancies within the standard cosmological model present profound challenges to our current comprehension of the Universe, necessitating a reevaluation of the foundational assumptions underlying the $\Lambda$CDM model. In this study, we propose a novel cosmological model to describe the expansion history of the Universe, characterized by late-time accelerated expansion driven by a 2-form field interacting with cold dark matter (CDM). Such fields inherently introduce a preferred direction in cosmic space, thereby deviating from the isotropy assumed in the FLRW metric and favoring the Bianchi geometries. For this preliminary exploration, we employed a locally rotationally symmetric (LRS) Bianchi I spacetime to investigate the existence of a
anisotropic accelerated attractors.

Unless forbidden by unknown symmetries, interaction between the dark sectors is plausible. We have considered a scenario wherein CDM and dark energy (DE) can exchange momentum. Motivated by coupling mechanisms arising from string compactifications, we have adopted an exponential coupling between these components, delineating the interaction kernel $\mathcal{Q}$ from a field-theoretical standpoint. This formulation enables us to express the Lagrangian for interacting CDM in terms of its density, consistent with its characterization as a pressureless perfect fluid.

Using dynamical systems analysis, we have mapped the asymptotic behavior of our model to two distinct attractors delineated by a bifurcation curve. Both attractors represent de-Sitter type solutions where the Universe undergoes perpetual expansion at a constant rate. However, these solutions diverge in their isotropic properties: one (DE1) preserves isotropy, while the other (DE2) promotes anisotropic expansion.

Of particular interest is the \emph{anisotropic de-Sitter solution} predicted by our model, which opposes existing literature assertions, such as those presented by Kaloper, who demonstrated that a the Kalb-Ramnod sector along a positive cosmological constant on top of certain Bianchi geometries violate the cosmic no-hair conjecture, like the diagonal Bianchi type-II. But in the simplest case of the Bianchi I spacetime, this conclusion does not hold. However, our findings suggest that replacing the cosmological constant $\Lambda$ with a potential encoding auto-interactions of the Kalb-Ramond 2-form field within a Bianchi I framework leads to viable anisotropic attractors. Future research will aim to extend these findings across a broader range of Bianchi geometries, varying potentials, and more complex configurations of the 2-form field, generalizing the so-called ``wonderland'' solutions for general $p$-forms.

Numerical integration of the autonomous equations, beginning from initial conditions deep in the radiation-dominated era and assuming an initially smooth Universe, have verified that our model mimics the expansion history of the $\Lambda$CDM model while predicting a significant late-time anisotropy sourced by DE. Indeed, it is this interplay between the 2-form field and the anisotropy which which enables the preservation of each field's respective dynamical properties. Our findings indicate that within the isotropic attractor, the 2-form field decays over time, and its associated potential approaches a constant value, effectively mimicking the behavior of a cosmological constant. Conversely, in the anisotropic attractor (DE2), the 2-form field remains dynamically active, with its persistence being directly sustained by the prevailing anisotropy. This distinction underscores the critical role of anisotropy in modulating the evolutionary pathways of fields within our model.

Isotropy violation in cosmological models is often linked to the presence of vector fields, making the distinction between 1-form (vector fields) and 2-form fields critical in understanding their respective influences on the expansionary dynamics. Contrary to the dynamical relation between the 2-form field and the anisotropy, our analysis reveals that the dynamics of the 1-form field $A_\mu$ is suppressed by the presence of the shear. This suppression leads to a scenario where the system converges to an isotropic attractor, suggesting that vector fields alone are insufficient to support a prolonged anisotropic accelerated expansion. However, it is important to emphasize that these observations are not general claims, as the dynamics might differ significantly under more general Bianchi metrics. We will delve into this important question in a future work.

We have analyzed the contribution of anisotropic DE to the CMB quadrupole. Our findings indicate that both the isotropic and anisotropic attractors are capable of maintaining a sufficient degree of anisotropy at present to potentially mitigate the discrepancies observed in the CMB quadrupole measurements. Hence, despite the Universe's asymptotic trend towards isotropic accelerated expansion, as in the (DE1) attractor, our results suggest that a significant level of anisotropy, which remains within observational constraints, can be relevant.

Our analysis reveals that interactions within the dark sectors, despite not altering the asymptotic behavior of the cosmological model, significantly influence its evolutionary dynamics. Specifically, we observe that the shear experiences an enhancement at high redshifts ($z_r < 100$), which could potentially affect the process of structure formation. Furthermore, the sign of the coupling strength, denoted as $\mu$, plays a critical role in these dynamics. We have determined that a positive $\mu$, indicating a decay of dark energy (DE) into cold dark matter (CDM), acts to suppress anisotropy. Conversely, a negative $\mu$, representing the decay of CDM into DE, leads to an increase in anisotropy.

We have used recent observational background data to constrain the parameter space of our model. Remarkably, this anisotropy is tightly constrained to approximately $\Sigma_0 \approx 10^{-4}$. This precision in constraining $\Sigma_0$ indicates a significant sensitivity of the dataset to the anisotropic aspects of the cosmological model. However, the evaluation of our model using the Akaike Information Criterion penalizes our model for the inclusion of two additional parameters, $\lambda$ and $\mu$, statistically favoring $\Lambda$CDM over our proposal. On the other hand, the observational data employed in our study does not sufficiently probe the sensitivities required to constrain the model parameters $\lambda$ and $\mu$. As such, these parameters remain largely unrestricted from the current dataset. However, we hope that observational data related to linear-order phenomena will provide invaluable insights for constraining the parameters of our model. Nevertheless, despite extensive exploration of linear perturbation theory in the context of Bianchi cosmologies in previous studies, we still lack of an accurate numerical implementation of it in a Boltzmann code. Consequently, a thorough and rigorous evaluation of anisotropic models continues to be beyond our current capabilities.

\section*{Acknowledgements}

This work was supported by Patrimonio Autónomo - Fondo Nacional de Financiamiento  para  la  Ciencia,  la  Tecnología  y  la  Innovación  Francisco  José  de  Caldas  (MINCIENCIAS - COLOMBIA)  Grant  No.   110685269447  RC-80740-465-2020,  projects  69723 and 69553 and by Vicerrectoría de Investigaciones - Universidad del Valle Grant No. 71357.

\bibliography{Bibli.bib} 

\end{document}